\documentclass[aps,prd,nofootinbib,showpacs,preprint,utf8]{revtex4-1}

\usepackage{amstext,amssymb}
\usepackage{amsmath}
\usepackage{tikz}
\usepackage{hyperref}
\usepackage{xspace}
\usepackage{color}
\usepackage{units}
\usepackage{slashed} 
\usepackage{braket}
\usepackage{footmisc}

\usepackage{wrapfig}
\definecolor{light-gray}{gray}{0.95}
\usepackage{tcolorbox}
\newcommand{\blue}[1]{{\color{blue} #1 }}

\begin{document}

\title{Minimal framework for asymmetric dark matter}
\title{Cogenesis of visible and dark sector asymmetry in a minimal seesaw 
framework}

\author{Utkarsh Patel}
 \email{utkarshp@iitbhilai.ac.in}
 \affiliation{Department of Physics, Indian Institute of Technology Bhilai, 
India}
 
 \author{Lekhika Malhotra}
 \email{lekhikamalhotra97@gmail.com}
 \affiliation{Department of Physics, Indian Institute of Technology Bombay, 
India}
 
 \author{Sudhanwa Patra}
 \email{sudhanwa@iitbhilai.ac.in}
 \affiliation{Department of Physics, Indian Institute of Technology Bhilai, 
India}
 
 \author{Urjit A.Yajnik}
 \email{yajnik@iitb.ac.in}
 \affiliation{Department of Physics, Indian Institute of Technology Bombay, 
India}

\date{\today}

\begin{abstract}
Recently there is a renewed interest in exploring the Dark sector of the universe in a more constrained way. Particularly in \cite{Datta:2021elq}, the FIMP ( Feebly Interacting Massive Particle) scenario was shown to be realized with a minimal extension of the SM with three
sterile neutrinos in the spirit of $\nu$MSM. In this paper, we show that without invoking any additional symmetries of the model, we can realize the idea of ADM (Asymmetric Dark Matter) signaling a common origin of the matter-anti-matter asymmetries in visible as well as Dark sectors. The model allows for a range of dark matter masses, $\sim 1 - 100$keV,
with correct active neutrino masses through the Type-I seesaw mechanism. Thus, the model explains the neutrino masses, Dark Matter abundance and replicates matter-anti-matter asymmetry of the visible sector in the Dark sector, all in a version of the  $\nu$MSM.  
The asymmetry in the dark sector is manifested in the predominance of one parity of the heavy neutrinos in the comoving frame, which should be determinable in the ongoing experiments.
\end{abstract}

\maketitle

\section{Introduction}
\label{sec:intro}
In the context of current theoretical high-energy physics development, the problems of tiny neutrino masses\cite{Super-Kamiokande:1998kpq, SNO:2002tuh, K2K:2002icj}, dark matter\cite{Julian:1967zz,SDSS:2003eyi} and matter-antimatter asymmetry\cite{Planck:2018vyg} could be considered as few of the most important hints for beyond the Standard Model Physics. Thus, a plausible extension of the Standard Model(SM) connecting the three would be highly desirable. With this motivation, in this work, we introduce three generations of right-handed neutrinos$(N_i, i=1,2,3)$, to account for the neutrino masses. Further, on the one hand, the lightest and stable right-handed  neutrino$(N_1)$ can be a Dark Matter candidate, in the spirit of the $\nu$MSM \cite{Asaka:2005pn}\cite{Asaka:2005an}\cite{Sahu:2005fe}\cite{Canetti:2012kh}. On the other hand, the out-of-equilibrium decay of the same species in the early universe can naturally explain the baryon asymmetry \cite{Yoshimura:1978ex}\cite{Weinberg:1979bt} via Leptogenesis\cite{Fukugita:1986hr}\cite{Luty:1992un} using sphaleron process\cite{Kuzmin:1985mm}.  

The goal of this paper is to extend the possibility of such scenarios to accommodate
 the asymmetric dark matter (ADM) scenarios. The energy densities of visible matter(VM) and dark matter(DM) in the Universe are of the same order\cite{WMAP:2003elm,Planck:2015fie}, with $\Omega_{\text{DM}}/\Omega_{\text{VM}}\sim5$. This serves as a motivation to look for common origins of both sectors in cosmological evolution. Observationally the VM asymmetry is \cite{SDSS:2003eyi,Planck:2018vyg} of the order of $10^{-10}$. Thus, a common origin is expected to leave an imprint of this asymmetry also in the dark sector. This is the basis for the framework of Asymmetric Dark Matter \cite{Nussinov:1985xr,Kaplan:1991ah}. Considerable subsequent developments for the ADM scenario can be found in \cite{Foot:2013msa,Blinnikov:1982eh,Berezhiani:2003wj,Ciarcelluti:2004ip,Hooper:2004dc,Kaplan:2009ag,Haba:2011uz,Dutta:2006pt,Shelton:2010ta,Falkowski:2011xh,vonHarling:2012yn,Bell:2011tn,Kuzmin:1996he,Dulaney:2010dj,
An:2009vq,Farrar:2005zd}. 

This work is along the lines of references ~\cite{Datta:2021elq},\cite{Bhattacharya:2021jli}, but differs from them
in terms of the dark matter candidate properties. In reference \cite{Datta:2021elq}, the authors present a minimal framework where the origins of neutrino mass, Dark matter and matter-antimatter asymmetry are explained in a unified way. The significance of their work lies in the fact that they have connected the small coupling required for FIMP 
realization with the smallness of the lightest active neutrino mass $m_1$ . Such a connection provides a stable FIMP dark matter in the keV mass range. Motivated by their idea, here we check whether such a FIMP scenario can also accommodate ADM possibility.
We find that with the addition of a singlet scalar ($\phi$) to their model, the Asymmetric FIMP scenario can be easily realised. Importantly, we refrain from invoking any new ad-hoc symmetry to stabilize the DM candidate. The DM stability comes out naturally within the construct of our framework. Such an asymmetric DM
has been discussed thoroughly in \cite{Falkowski:2011xh}, and we make use of their ideas for the evolution of asymmetries in our model as well. We also discuss the possibility of detecting such origin of asymmetry in the Dark sector in ongoing experiments.

The outline of this work is as follows. In section~\ref{sec:typeI}, we briefly present the Type-I seesaw mechanism responsible for neutrino mass generation. Section~\ref{sec:frame} describes the specifics of our model, including the flavor mixing through Yukawa couplings for both sectors. In section~\ref{sec:asymm}, an analytical expression for the asymmetric decay of $N_2$ into the visible and dark sectors is derived, considering the complex value of couplings. In the next section~\ref{sec:adm}, we obtain quantitative results for the cosmological evolution of $N_2$ and asymmetries in both sectors by 
numerically solving the Boltzmann equations. All the Detection constraints have been discussed in section~\ref{sec:det}. Finally, in~\ref{sec:con}, we discuss the outcomes and future scope of our work. All the other relevant details and calculations can be found in the appendix~\ref{app:il}. 

\begin{figure*}
\includegraphics[scale=0.6]{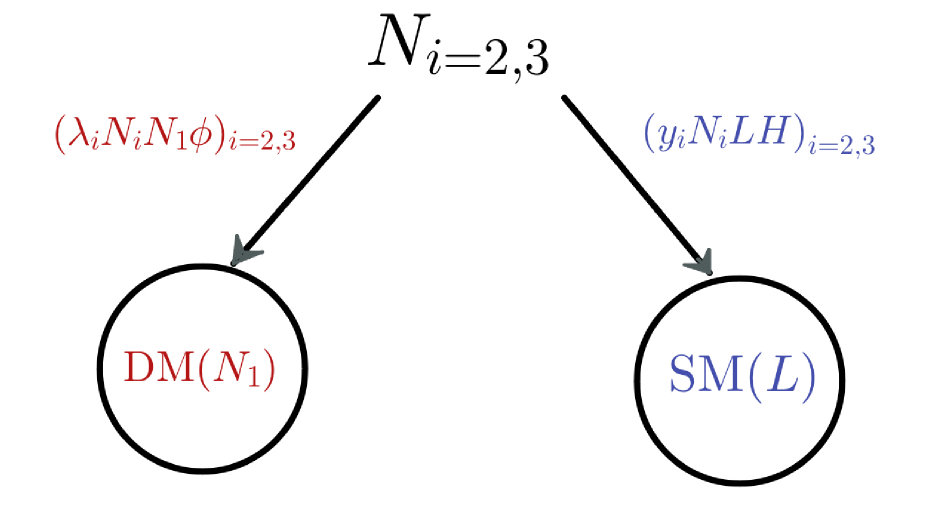}
\caption{Schematic of our model framework. Here, $N_1$ is taken as a DM 
candidate, and $L,H$ refer to the usual Leptons and Higgs doublet of 
SM, respectively and $\phi$ is the extra scalar singlet.}
\label{fig:1}
\end{figure*}

\section{Neutrinos in Type-I seesaw}
\label{sec:typeI}

The type-I seesaw mechanism for light neutrino masses, popularly known as canonical seesaw, was originally proposed by \cite{Minkowski:1977sc, 
Mohapatra:1979ia, Yanagida:1979as, Schechter:1980gr}. It is the simplest 
mechanism that explains the lightness of SM neutrino masses with the 
introduction of heavy right-handed neutrinos. Here we recapitulate the essentials for reference. The right-handed neutrinos being singlet under SM, are used to write down a Dirac mass term between SM neutrinos 
$\nu_L$ and right-handed neutrinos $N_R$ along with a bare Majorana mass term for right-handed neutrinos $N_R$. The relevant mass terms for the type-I seesaw scheme are given by
\begin{eqnarray}
\mathcal{L}_{\rm Type-I} &=& \mathcal{L}_{M_D}+\mathcal{L}_{M_R} 
\nonumber \\
\mathcal{L}_{M_D} &=& - \sum_{\alpha, \beta} \overline{\nu_{L \alpha}} 
[M_D]_{\alpha k} N_{Rk} \mbox{+ h.c.}\nonumber \\
\mathcal{L}_{M_{R}} &=& - \frac{1}{2} \sum_{j, k} \overline{N^c_{Rj}} 
[M_{R}]_{jk} N_{Rk} \mbox{+ h.c.}
\end{eqnarray}
The first term is the Dirac neutrino mass connecting both left- and right-handed 
chiral fermions, while the second term is the Majorana mass term involving only a 
singlet right-handed chiral fields. \begin{eqnarray}
 \mathcal{L}^M_N \supset -\frac{1}{2}\begin{pmatrix} \overline{(\nu_L)} & 
\overline{N_R^C}                                     \end{pmatrix}
 \begin{pmatrix}  0 & M_D \\ M_D^T & M_R	 \end{pmatrix}                                                                                                    
\begin{pmatrix} \nu_L^C \\ N_R
\end{pmatrix}
\end{eqnarray}
with, Dirac mass matrix $(M_D)_{I\alpha}=y_{I\alpha}v/\sqrt{2}$ where 
$Y_{I\alpha}$ is given in Eq.(\ref{eq:mnu_MR_def}). The structure of complex 
Yukawa matrix can be found by using Csasa-Ibara Parameterization. The weak or 
flavor eigenstates defined for the left-handed (LH) and right-handed(RH) 
neutrinos are as follows,
\begin{eqnarray}
\nu_{\alpha L} = \begin{pmatrix}
\nu_{eL} \\ \nu_{\mu L} \\ \nu_{\tau L}
\end{pmatrix}\, , \quad 
N_{Rk} = \begin{pmatrix}
N_{R1} \\ N_{R2} \\ N_{R3}
\end{pmatrix}\, .
\end{eqnarray}
Similarly, the mass eigenstates for LH and RH neutrinos are given by-
\begin{eqnarray}
\nu_{i} = \begin{pmatrix}
\nu_{1} \\ \nu_{2} \\ \nu_{3}
\end{pmatrix}\, , \quad 
N_{k} = \begin{pmatrix}
N_{1} \\ N_{2} \\ N_{3}
\end{pmatrix}\, .
\end{eqnarray}
The flavor eigenstates and mass eigenstates are related to each other as 
follows,
\begin{eqnarray}
 &&\nu_{\alpha }= U_{\alpha i} \nu_{i}+S_{\alpha j} N^c_{j} \nonumber \\
 &&N_{\beta }=T^*_{\beta k} \nu^c_{k}+V^*_{\beta l}N_{l}
\end{eqnarray}
 
After a change of basis, the resulting mass matrix for neutrinos is given by,
\begin{equation}
 M_\nu \equiv  \left[ 
\begin{array}{c | c} 
\blue{\bf 0} &  M_{D}
\\ 
\hline 
  M^T_{D} & \blue{M_R}\\
 \end{array} 
\right]
\end{equation}
is diagonalized by the unitary matrix
\begin{equation}
 \mathcal{V} \equiv \begin{pmatrix}
           U & S \\ T & V
          \end{pmatrix}
          \label{eq:appA1}
\end{equation}
The unitarity conditions $\mathcal{V}^\dagger \mathcal{V} = \mathbb{I}_{6\times 
6} = \mathcal{V} \mathcal{V}^\dagger$ further yield
\begin{eqnarray}
&&U T^\dagger + S V^\dagger = 0 = S^\dagger U + V^\dagger T \nonumber \\
&&U U^\dagger + S S^\dagger = 1 = T T^\dagger + V V^\dagger \,.
\label{eq:diagonalization}
\end{eqnarray}
Within the diagonal basis of charged leptons and heavy Majorana RH-neutrinos, 
the light neutrino mixing matrix $V_\nu \equiv U_{\rm PMNS}$ and $V_R \equiv 1$. 
As a result, the resulting mixing matrices $U$, $S$, $T$ and $V$ can be expanded 
as
\begin{equation}
  U \equiv \left[\mathbb{I}_{3\times 3} - 
\frac{1}{2}M^{}_DM_R^{-1}(M^{}_DM_R^{-1})^\dagger\right] U_{\rm PMNS}, \quad 
\label{eq:matrixU}
 \end{equation}
 \begin{equation}
  V \equiv \left[\mathbb{I}_{3\times 3} - \frac{1}{2}(M^{}_DM_R^{-1})^\dagger 
M^{}_DM_R^{-1}\right] ,
\label{eq:matrixV}
\end{equation}
\begin{equation}
   S \equiv M^{}_DM_R^{-1} , \quad \label{eq:matrixS}
  \end{equation}
  \begin{equation}
    T \equiv -(M_D M_R^{-1})^\dagger U_{\rm PMNS}\, , \label{eq:matrixT} 
\end{equation}

The simplified mass relations~\cite{Branco:2020yvs} for active light neutrinos and sterile neutrinos 
is given by
\begin{eqnarray}
&& m_\nu =  - M_DM_R^{-1}M_D^T  \approx \ -\frac{M_D^2}{M_R}, \nonumber \\[1mm]  
&& M_{D}=-\iota UD_{\sqrt{m_\nu}}R^TD_{\sqrt{M_R}}= y_{\nu}v/\sqrt{2}.
\label{eq:mnu_MR_def}
\end{eqnarray}
 Here, $m_i$ ($M_i$) are the light (heavy) neutrino mass eigenvalues, $y_{\nu}$ is complex Yukawa matrix, $R$ is a complex orthogonal matrix,
and $D_{\sqrt{m_\nu}} \equiv {\rm 
diag}(\sqrt{m_1},\sqrt{m_2},\sqrt{m_3})$ while $D_{\sqrt{M_R}} \equiv {\rm 
diag}(\sqrt{M_1},\sqrt{M_2},\sqrt{M_3})$.


\section{Theoretical Framework}
\label{sec:frame}
For implementing neutrino mass, leptogenesis, and dark matter asymmetry in a 
unified framework, let us consider the most straightforward extension of the SM 
with one copy of right-handed neutrino $(N_i, i=1,2,3)$ per generation 
and one extra scalar singlet $\phi$. The overall perspective of the framework
is schematically given in Fig.~\ref{fig:1}. The Lagrangian for 
the model is taken to be 
\begin{eqnarray}
&&\mathcal{L}_{} = \mathcal{L}_{SM}+\mathcal{L}_{N_R}+\mathcal{L}_{\phi}
\nonumber \\
&&\mathcal{L}_{N_R} = i\,\overline{N_{Rk}}\, \slashed{\partial}\, N_{Rk} 
-\sum_{\alpha, k} y_{\alpha k} \overline{\ell}_{\alpha L}\, \widetilde{H}\,
N_{Rk} 
\nonumber \\&& \hspace*{1cm}
- \frac{1}{2} \overline{N^c_{Rj}} M_{jk} N_{Rk}-\lambda_{jk}\phi 
\overline{N_{Rj}^c}N_{Rk}
\mbox{+ h.c.}
\nonumber \\
&&\mathcal{L}_{\phi} =m_{\phi}^2 \phi^{2} + 
\frac{\lambda_{\phi}}{2}|\phi|^4 + \lambda_{\phi 
H}\phi^2(H^{\dagger}H)
\nonumber \\&& \hspace*{1cm}
+\kappa_{\phi}(H^{\dagger}H)\phi\mbox{ + h.c.}
\label{eq:tf1}
\end{eqnarray}

In the $\mathcal{L}_{SM}$ part, not fully displayed, we use $\ell_L 
\equiv \big(\nu_L, \ell_L \big)^T$ and $H\equiv \big(\phi^0, \phi \big)^T$ as 
the SM lepton doublet and Higgs doublet, the conjugate Higgs field is defined as 
$\widetilde{H} = i \tau_2 H^*$. And our primary interest will be the 
term $h_{\alpha \beta} \overline{\ell}_{L\alpha} H \ell_{R\beta}$ 
being the SM Yukawa coupling terms for the leptons. 
We select a basis in which the $h$-matrix is flavor diagonal, i.e., 
$h_{\alpha \beta} \overline{\ell}_{L\alpha} H \ell_{R\beta} \equiv h_{\alpha 
\alpha} \overline{\ell}_{L\alpha} H \ell_{R\alpha}$. 
The $N_{Rk}$ is a basis of right-handed Weyl spinors, which after due mixing 
of neutrino, states give the physical Majorana fields $N\equiv N^c$. 
The $N_{Ri}$ are chosen in a way that the matrix $M_{j k}$ is diagonal.
Thus the matrix $y_{\alpha k}$ remains the source of flavor mixing.
The eigenvalues of $M$ are assumed to be large to accord with the Type-I 
seesaw mechanism.

We have assumed a hierarchical RH-neutrino spectrum 
such that the lightest RH-neutrino becomes a dark matter candidate while the 
CP-violating and out-of-equilibrium decays of the other two heavy Majorana 
RH-neutrinos explain the matter-antimatter asymmetry of the universe via thermal 
leptogenesis~\cite{Fukugita:1986hr}. The important feature of the model is that 
after rewriting the Lagrangian in the mass eigenstates and extracting the 
interactions of RH-neutrinos with $W,Z$ boson, Higgs $H$ and leptons $\ell$, it 
is possible to have decays of $N_{2,3} \to \ell H$ in visible sector and new 
decay channels $N_{2,3} \to N_1 + H, N_1 +Z,N_1+\phi$ in the dark-sector leading 
to asymmetries in both sectors simultaneously. Thus, light neutrino mass 
generation, lepton asymmetry, and dark matter asymmetries are simultaneously 
explained in this next-to-minimal framework of the canonical seesaw scheme.

\begin{figure*}
\includegraphics[scale=0.9]{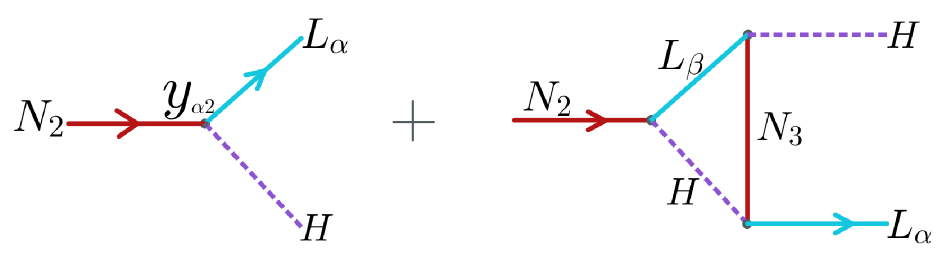}
\caption{Feynman Diagram corresponding to the tree level and loop level 
interaction of RHNs and Higgs. These interactions are responsible for the 
Lepton asymmetry generation through CP-violating complex phases. 
}
\label{fig:scat2}
\end{figure*}

\section{Asymmetry Generation}
\label{sec:asymm}
Dynamical generation of  excess matter over anti-matter (also called Baryogenesis, or abbreviated BAU),
in the early universe is an appealing challenge 
to BSM particle physics. 
The salient ingredients required for this to unfold were identified by Sakharov~\cite{Sakharov:1967dj}. The three "Sakharov conditions" relevant to us are (1)  Lepton number violation 
 (2) CP violation
 (3) Out-of-equilibrium dynamics.
Several approaches have been discussed in the literature to explain 
Baryogenesis~\cite{Fukugita:1986hr,Affleck:1984fy,Luty:1992un} . 
A fruitful approach toward this realization in the context of neutrinos is the possibility
of lepton number violation caused by the decay of Majorana neutrinos.
Decay of these heavy RH neutrinos whose mixing matrix carries complex phases 
leads to Leptogenesis~\cite{Fukugita:1986hr}\cite{Davidson:2008bu} in the early universe.
 This excess lepton number thus generated transforms partially into baryon number 
excess via sphaleron processes~\cite{Kuzmin:1985mm}.  
In our work, we've used the thermal leptogenesis~\cite{Buchmuller:2004tu} 
to explain the BAU in the visible sector and the handedness asymmetry in the dark sector.

The interaction lagrangian for RH neutrinos and SM particles can be written in 
flavor basis as well as in a mass basis. The detailed calculation for which are 
given in the appendix~\ref{app:il}.
\begin{itemize}
 \item The charged current interaction involving  RH-neutrino $N_j$ and $W$ 
boson is given as,
\begin{eqnarray}
&&\mathcal{L}_{\rm CC} = - \frac{g}{\sqrt{2}} \sum_{i,j=1}^{3} 
 \overline{\ell}_\alpha S_{\alpha j} \gamma^{\mu} P_L \nu_\alpha  W_{\mu}^- + \mbox{h.c.}
\quad \mbox{[flavor basis]}\nonumber \\
&& \subset - \frac{g}{\sqrt{2}} \sum_{\alpha,j=1}^{3} \overline{\ell}_j 
S_{ij}\gamma^{\mu} P_L N_j W_{\mu}^- + \mbox{h.c.} \quad \mbox{[mass 
basis]} \nonumber
 \label{eq:asymm1}
\end{eqnarray}
Here, three generations of charged lepton are represented by $\ell_i$ and $P_L 
=\frac{1}{2} \big(1-\gamma_5 \big)$ is the projection operator. 
\item In the same way, the neutral current interactions involving RH-neutrino 
$N_j$ and $Z$ boson are given as,
\begin{eqnarray}
&&\mathcal{L}_{\rm NC} \subset - \frac{g}{\cos \theta_W} Z_\mu \bigg[
\sum_{k,j=1}^{3} \overline{N}_k^c \gamma^\mu P_L \big(S^\dagger S\big)_{kj} N_j
\nonumber \\
&&\hspace*{1.9cm}
+ \bigg\{\overline{\nu}_k \gamma^\mu P_L \big(U^\dagger S\big)_{kj} N_j+ 
\mbox{h.c.} \bigg\}
\bigg]
 \label{eq:asymm2}
\end{eqnarray}
with $\cos\theta_W$ as the cosine of the known Weinberg mixing angle.
\item The most important interactions of RH-neutrinos $N_j$ with the SM Higgs 
and the scalar singlet is given as,
\begin{eqnarray}
\begin{split}
&&\mathcal{L}^{N}_{H}\subset \frac{\sqrt{2}}{v} H \sum_{i,j=1}^{3}\Big[ 
\bar{\nu}_{i}(U^{\dagger}S)_{ij}M_j N_{j}
+ \overline{N}^c_{i}(S^{\dagger}S)_{ij}M_j N_{j}\Big] \\
&& + h.c.,\label{ym} 
 \end{split}
 \label{eq:asymm3}
\end{eqnarray}
\begin{eqnarray}
 \mathcal{L}_s^N \supset \lambda_{jk}\phi N_{Rj}N_{Rk}
 \label{eq:asymm4}
\end{eqnarray}
\end{itemize}
The first term $\frac{\sqrt{2}}{v} H \overline{\nu}_{i} 
\big(U^{\dagger}S\big)_{ij}\, M_j N_{j}$ (or, equivalent interaction in flavor 
basis, $Y_{\alpha j} \overline{\nu}_\alpha \widetilde{H} N_{Rj}$) in 
Eq.(\ref{eq:asymm3}) is the relevant term for thermal leptogenesis creating 
CP-asymmetry in the lepton sector, which is further converted into baryon asymmetry. 
The out of equilibrium decay of $N_{2,3} \to \nu + h$ can generate required lepton 
asymmetry. The second term  $\frac{g }{2 m_W} h \sum_{i,j=1}^{3}  
\overline{N}^c_{i}(S^{\dagger}S)_{ij}M_j N_{j}$ in Eq.(\ref{eq:asymm3}) mediating 
decay of $N_{2,3} \to N_1 + H$ and its CP-violation comes from complex Yukawa 
couplings. Here, S is defined in Eq(~\ref{eq:matrixS}). This term generates 
CP-asymmetry in the dark sector, but the generated asymmetry is not sufficient 
for the asymmetries in both sectors to co-evolve. Thus with the help of a 
singlet scalar $\phi$, mediating interactions $(N_{2,3}\rightarrow N_1+\phi)$ as 
in Eq.(\ref{eq:asymm4}), a significant asymmetry in the dark sector is restored.
\subsection{CP violations in Visible Sector}
\label{subsec:cpvis}
The out-of-equilibrium decays of the heavy Majorana neutrinos involve
CP-asymmetric amplitudes if the neutrino Yukawa couplings $y_{k\alpha}(k=2,3)$ are 
complex. This CP asymmetry afflicts the conservation of the Lepton charges 
$L_{\alpha}(\alpha=e,\mu,\tau)$, the asymmetry $(\epsilon_{k\alpha})$ being defined as
\begin{equation}
 (\epsilon_{k\alpha})=\frac{\Gamma(N_{k}\rightarrow 
l_{\alpha}H)-\overline{\Gamma}(N_{k}\rightarrow 
\overline{l}_{\alpha}\overline{H})}{\Gamma_{Dk}}
\label{eq:cplep1} 
\end{equation}
Here, $\Gamma_{Dk}$ is the total decay rate for $N_{k}$ in the visible sector 
and is given as
\begin{equation}
\begin{split}
\Gamma_{Dk} & =\sum\limits_{\alpha}[\Gamma(N_{k}\rightarrow 
l_{\alpha}H)+\overline{\Gamma}(N_{k}\rightarrow 
\overline{l}_{\alpha}\overline{H})]\\
&=\frac{[yy^{\dagger}]_{kk}M_{k}}{8\pi}
\end{split}
\label{eq:cplep2} 
\end{equation}

We work with the assumption of hierarchy $M_1<<M_2<<M_3$, so that it is the decays
of the $N_2$ that essentially determines the sought-after asymmetries.
The expression for $CP$ asymmetry can be obtained by putting $k=2$ in 
Eq.(\ref{eq:cplep2}).
\begin{figure*}
\includegraphics[scale=1]{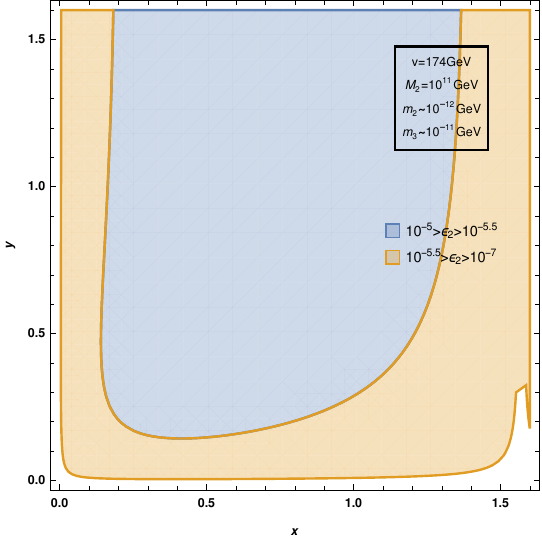}
\caption{Figure shows a scatter plot for the allowed values of real and imaginary parts
$x$ and $y$ of the complex rotation angle in the $R$ matrix, Eq.(\ref{eq:mat-1}) 
for various value 
ranges of asymmetry parameter in visible sector$(\epsilon_l)$ and  
values of active neutrino masses $m_2$ and $m_3$.The mass of $N_2$ 
is taken to be $10^{11}\text{ GeV }$. The values of $x$ and $y$ are varied  between 0 to 
1.6.}
\label{fig:scat1}
\end{figure*}

The Feynman diagrams corresponding to the decay of $N_2$ into $l$ at the tree 
and loop level that contribute to the CP asymmetry in the visible sector 
$(\epsilon_l)$ is given in Fig.~\ref{fig:scat2}. The analytical expressions 
for $\Gamma\text{'s}$ corresponding to various decay modes can be found in 
appendix~\ref{app:dr}. The expression for flavor asymmetry is given by
\begin{equation}
\begin{split}
& \epsilon_{l\alpha}= -\frac{3}{16\pi(h^\dagger h)_{ii}}\sum\limits_{j\neq i}\{ 
\text{Im} [h^*_{\alpha i}h_{\alpha j} (h^{\dagger}h)_{ij}] 
\frac{\xi(x_j/x_i)}{\sqrt{x_j/x_i}} \\ 
&+\frac{2}{3(x_j/x_i-1)}\text{Im}[h^*_{\alpha i}h_{\alpha 
j}(h^{\dagger}h)_{ji}]\}
\end{split}
\label{eq:cplep4} 
\end{equation}
Here, $x_j\equiv M_j^2/M_1^2$ , $h$ is a general coupling variable, and the loop 
function $\xi(x)$ as derived in reference\cite{Blanchet:2008hg}is given by
\begin{equation}
\xi(x)=\frac{2}{3} x\bigg[(1+x)\text{ln} \bigg( 
\frac{1+x}{x}\bigg)-\frac{2-x}{1-x} \bigg]
\label{eq:cplep5}
\end{equation}
for ${x_j\gg x_i}$. Summing the value for total CP asymmetry associated to the 
decays of $N_2$ over all the lepton flavors is given as~\cite{Hagedorn:2017wjy}:
\begin{equation}
\begin{split}
& \epsilon_l \equiv \sum\limits_{\alpha} \epsilon_{l\alpha}\\
& = \frac{3}{16\pi}\frac{M_2}{M_3}\frac{1}{[y y^{\dagger}]_{22}} \text{Im} \{ 
(y^{\dagger}y)_{23}^2 \}
\end{split}
\label{eq:cplep9}
\end{equation}
Here, $(y_\nu)_{ij}$ denotes the Yukawa coupling matrix for the visible sector, 
whose expression defined in Eq(~\ref{eq:mnu_MR_def},\ref{eq:Yukawa_coupling}) and using this value in 
Eq.(\ref{eq:cplep9}), we get
\begin{equation}
\epsilon_l= -\frac{3}{16\pi v^2} 
\frac{M_2.\text{Im}(\sum\limits_{j}m_j^2R_{2j}^2)-M_1.\text{Im}(\sum\limits_{j}
m_j^2R_{j2}^2)}{\sum\limits_{i}m_i|R_{2i}|^2}
\label{eq:cplep6}
\end{equation}
With the assumed hierarchy $M_1<<M_2<<M_3$, the above expression can be approximated as:
\begin{equation}
\epsilon_l \simeq -\frac{3}{16\pi v^2} 
\frac{M_2.\text{Im}(\sum\limits_{j}m_j^2R_{2j}^2)}{\sum\limits_{i}m_i|R_{2i}|^2}
\label{eq:cplep7}
\end{equation}
The structure of the complex orthogonal matrix $R$ is taken to be
\begin{equation}R= \begin{bmatrix} 1 & 0 & 0\\
0 & \cos w & \sin w\\
0 & -\sin w & \cos w
\label{eq:mat-1}
\end{bmatrix}
\end{equation}
with $w=x+\iota y$ being a complex rotation angle.

Eq.(\ref{eq:cplep9}) can also be written simply in terms of mixing 
matrices~\ref{eq:matrixU} , \ref{eq:matrixS} as:
\begin{equation}
 \epsilon_l=\frac{3}{16\pi}M_2M_3{\bigg(\frac{g}{2M_w}\bigg)}^2\frac{\text{Im}[{(S^{\dagger}S)}_{23}^2]}{{[SS^{\dagger}]}_{22}}
 \label{eq:cplep10}
\end{equation}
using $U$ matrix as ${I}$.

Inserting the values of masses$(m_2, m_3 \text{ and } M_2)$ and higgs vacuum 
expectation value$(v)$ in Eq.(\ref{eq:cplep7}), a scatterplot of $\epsilon_l$ as a 
function of angle variables $x$ and $y$ can be plotted as in Fig.~\ref{fig:scat1}.

\begin{figure*}
\includegraphics[scale=0.9]{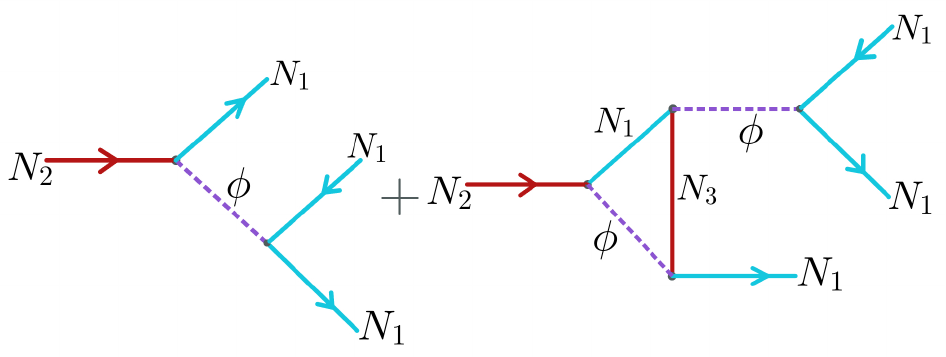}
\caption{Feynman Diagrams corresponding to the tree level and loop level 
decays of heavy RHNs into $N_1$ and singlet scalar $\phi$. These interactions are responsible 
for generation of relic dark matter $N_1$ with an asymmetry in handedness due to
the $CP$ violation.}
\label{fig:dark2}
\end{figure*}

\subsection{CP violations in Dark Sector}
\label{subsec:cpdark}
In the previous sub-section~\ref{subsec:cpvis}, we see from 
Fig.~\ref{fig:scat1} that asymmetry parameter in visible sector $\epsilon_l$ 
is of the order of $10^{-5}-10^{-8}$. Therefore, the asymmetry required in the dark sector must also be comparable in magnitude, given the fact that asymptotic abundance in the dark sector only depends on its initial number density and the asymmetry Eq.(\ref{eq:boltz1}). Keeping the framework minimal and not adding a singlet scalar to our model leaves the decay modes of $N_2$ into $N_1$ via $H$,$W$, and $Z$ only, which are insufficient to produce the desired asymmetry in the dark sector. Using Eq.(\ref{eq:cplep4}) to derive the asymmetry in the dark sector without adding a singlet scalar gives the expression for asymmetry 
$\epsilon_D$ as:

\begin{equation}
 \epsilon_{D(\text{H-channel})}=-\frac{1}{8\pi}\frac{1}{[\lambda'\lambda'^{
\dagger}]_{22}}\text{Im}[(\lambda'^{\dagger}\lambda')_{23}^2].f(x_j)
 \label{eq:cpdark01}
\end{equation}

Here, $\lambda'$ is the coupling for $N_2$ decay into $N_1$ via the Higgs channel. 
Using Eq.(\ref{eq:appA1} , \ref{eq:matrixS}), the final structure of 
Eq.(\ref{eq:cpdark01}) turns out to be:

\begin{equation}
 \begin{split}
\epsilon_{D(\text{H-channel})}&=-\frac{1}{8\pi}.\frac{g^4}{16M_{w}^4}
\bigg(-\frac{3}{2}\bigg)\frac{M_2}{M_3}\times\\
  &\frac{M_2^2.M_3^2\text{Im}[(S^{\dagger}SS^{\dagger}S)_{23}^2]}{\frac{M_2^2g^2}{4M_w^2}[S^{\dagger}SS^{\dagger}S]_{22}} \approx \mathcal{O}(10^{-22})
 \end{split}
 \label{eq:cpdark02}
\end{equation}

Comparing the above equation with Eq.(\ref{eq:cplep10}), we see that given the 
expressions for $S$ and $U$ matrices from Eq.(\ref{eq:matrixS} , \ref{eq:matrixU}), the value of 
$\epsilon_l$ will always be many orders greater in magnitude than the value of 
$\epsilon_D$. Hence, the Higgs channel is incapable of producing the needed 
asymmetry.

\par Thus, our model here considers an extra singlet scalar $\phi$ and sets 
lightest right-handed neutrino $N_1$ as the dark-matter candidate. The 
out-of-equilibrium decays of the other two heavy right-handed neutrinos 
$N_2, N_3$ can produce a  non-zero CP asymmetry in the dark sector, if the 
neutrino dark-yukawa couplings $\lambda_{k\alpha}(k=2,3)$ are complex. For 
simplicity, considering the case of only $N_2$ decay, this CP 
asymmetry$(\epsilon_{D})$ for dark-matter particles through the decay of the 
heavy right-handed neutrino $N_2$ via $\phi-$ channel, determines the evolution 
of Dark Matter asymmetry and is defined as:
\begin{equation}
 \epsilon_D=\frac{\Gamma(N_2\rightarrow 
N_1x)-\overline{\Gamma}N_{2}\rightarrow 
\overline{N}_{1}x)}{\Gamma_{D2}}
\label{eq:cpdark1}
\end{equation}
with $x= H,Z,\phi$.
Here, $\Gamma_{D2}$ is the total decay rate for $N_{2}$ in the dark sector and 
is given by:
\begin{equation}
\begin{split}
\Gamma_{D2} &=\Gamma(N_{2}\rightarrow 
N_{1}\phi)+\overline{\Gamma}(N_{2}\rightarrow \overline{N}_{1}x)\\
&=\frac{[\lambda\lambda^{\dagger}]_{22}M_{2}}{8\pi}
\end{split}
\label{eq:cpdark2} 
\end{equation}

The analytical expressions for $\Gamma\text{'s}$ corresponding to various decay 
modes can be found in appendix~\ref{app:dr} and the Feynman diagrams 
corresponding to the decay of $N_2$ into $N_1$ at the tree and loop level via 
higgs$(H)$, that contribute to the CP asymmetry in the dark sector$(\epsilon_D)$ 
is given in Fig.\ref{fig:dark2}.
Using Eq.(\ref{eq:cplep4}), we have
\begin{equation}
\epsilon_{D}= 
\frac{3}{16\pi}\frac{M_2}{M_3}\frac{1}{[\lambda\lambda^{\dagger}]_{22}}\text{Im} 
[(\lambda^{\dagger}\lambda)_{23}^2]  \approx \mathcal{O}(10^{-8})
\label{eq:cpdark3} 
\end{equation}
where, $\lambda$ is defined in Eq(~\ref{eq:Dark_coupling})
\begin{figure*}
\includegraphics[scale=0.51]{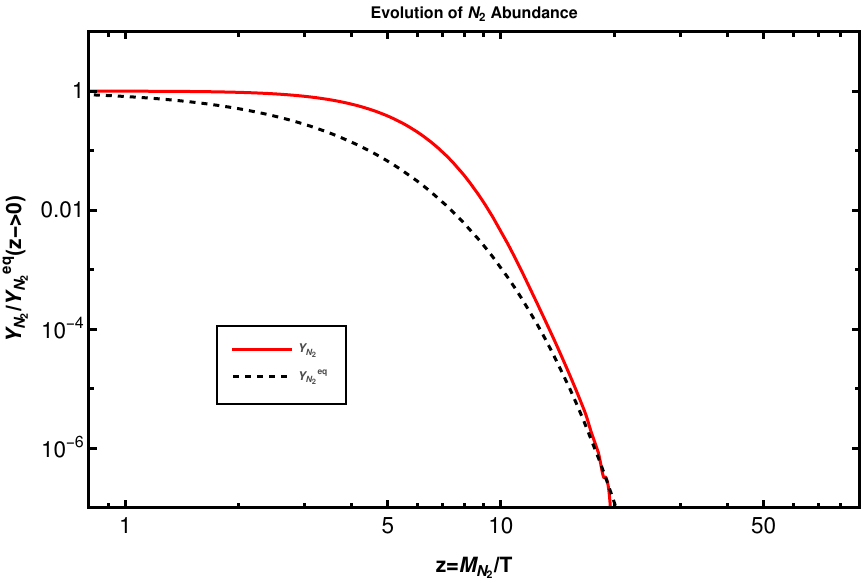}
\includegraphics[scale=0.52]{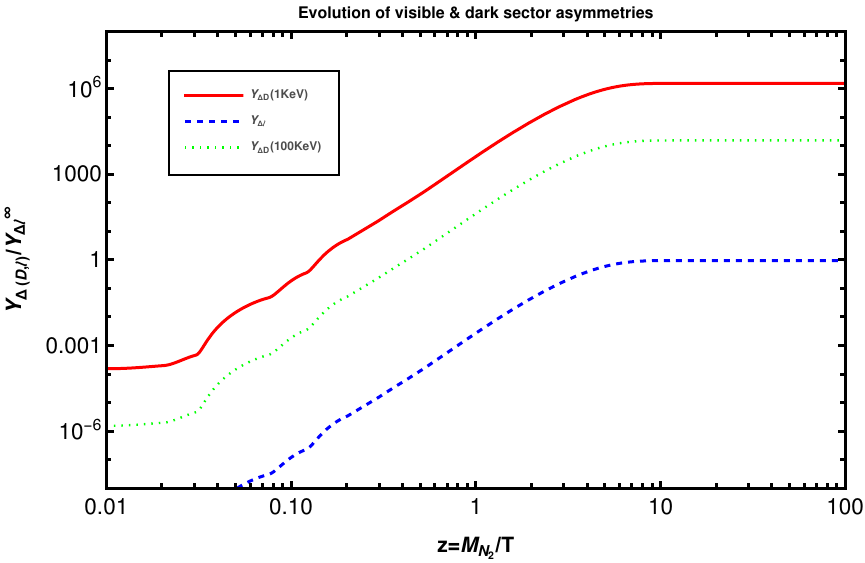}
\caption{The plots here show the co-evolution of the visible and dark sector 
asymmetries from the decay of heavy right-handed neutrinos after solving the 
Boltzmann Eq.(\ref{eq:boltz7}). On the left, the evolution of $N_2$ 
abundance( red line) is shown with respect to the equilibrium number density of 
$N_2$(black). On the right side, a plot is shown for the asymmetry abundance 
normalized to the asymptotic lepton abundance of $\Delta l$(blue dashed) and for 
$\Delta_{N1}$ with $m_{N1}=1 \text{ keV }$(red line) and $m_{N1}=100 \text{ keV 
}$(green dotted). The value of $M_{N_{2}}$ is taken to be $10^{11}\text{GeV}$.}
\end{figure*}

\section{Cosmological evolution of the asymmetry}
\label{sec:adm}
The generation of the Visible and Dark sector asymmetry occurs in the following 
steps, 
\begin{itemize}
\item A population of the heavy right-handed Majorana neutrinos, $N_{2,3}$, are  
generated in the early universe. 
\item At temperatures below $M_{N_{2,3}}$, these neutrinos decay out of 
equilibrium to both sectors (visible and dark sectors). 
The CP-violating decays lead to a lepton number asymmetry in both the SM and 
dark matter sector.
\item As the universe cools well below $M_{N_{2,3}}$, the washout of lepton 
asymmetry via inverse decay processes and its transfer between the 2 sectors, 
becomes inefficient, and the asymmetries are frozen-in.
The asymptotic asymmetry can, in general, be different in the two sectors due to 
different branching fractions and/or washout effects.  
\item The SM lepton asymmetry is transferred into baryon asymmetry via the well astonished
electroweak sphalerons.  
The symmetric baryon component is almost entirely wiped out by hadronic 
annihilations, and only the asymmetric component survives.  
\item 
In the Dark sector, the relic abundance is set by the major asymmetric DM component, and the DM receives a Majorana mass given as $M_{1} N^T_1 N_1$.
\end{itemize}
The asymptotic asymmetries at late times, as calculated through observational data, are expressed in terms of created asymmetry and washout parameters using~\cite{Iminniyaz:2011yp}as:
\begin{eqnarray}
Y_{\Delta L}^{\infty}&=&\epsilon_l\eta_2Y_{N2}^{eq}(0)\simeq 2.6\times10^{-10} 
\nonumber \\
Y_{\Delta\chi}^{\infty}&=&\epsilon_{D}\eta_{D}Y_{N2}^{eq}(0)\simeq 
2.6\times10^{-10}\bigg(\frac{\text{GeV}}{m_{\chi}}\bigg)
\label{eq:boltz1}
\end{eqnarray}
Here,$Y_{N2}^{eq}(0)$ is the initial equilibrium number density~\cite{Kolb:1990vq} of $N_2$ and is given  as $Y_{N2}^{eq}(0)=135\zeta(3)/4\pi^4g*$ with 
$g*\sim106$ being the total relativistic degrees of freedom at $T\sim M_{N_2}$. 
As shown in the next section, the washout efficiency parameters$(\eta)$ in both sectors 
turn out to be $\mathcal{O}(1)$ and imported in Eq.(\ref{eq:boltz1}).

\subsection{Boltzmann Equations}
\label{sec:boltz}
\begin{figure}
 \includegraphics[scale=0.8]{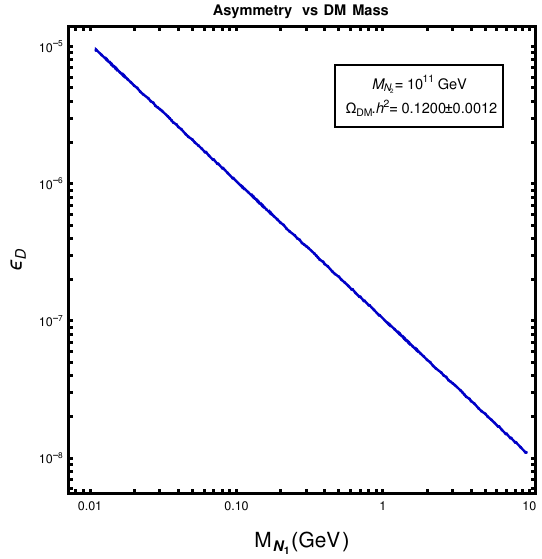}
 \caption{Plot shows the allowed values of Dark Matter candidate mass and the 
corresponding asymmetry in the dark sector required to obtain the correct relic 
density. The plot assumes the contribution to dark sector asymmetry only through 
the scalar mediator $\phi$ channel as contribution through other channels is 
comparatively very low.}
 \label{fig:scat3}
 \end{figure}

\begin{enumerate}
 \item The cosmological evolution of the heavy right-handed neutrinos and the 
asymmetries in both sectors are described by the Boltzmann Equations.
 \item We introduce the abundance yield $Y_{N_2} = n_{N_2}/s$ where $n_{N_2}$ is 
the number density of the particle $N_2$ and $s$ is the entropy density. Also, 
the evolution of equilibrium number density $Y_{N_2}^{eq}$ is important for the 
analysis, which is expressed as:
\begin{equation}
Y_{N_2}^{eq}=\frac{45}{2\pi^4g*}(0.9z^2K_n(z))  
\label{eq:boltz3}
\end{equation}
Here, $K_n(z)$ is the modified Bessel's function of the second kind.

 \item We are interested in the evolution of the asymmetries $Y_{\Delta l, 
N_{1}} = Y_{l,N_{1}} - Y_{\bar{l},\overline{ N_{1}}}$ as a function of time (or 
temperature $T$), assuming these asymmetries vanish at early times, given that 
$N_1$ is the Dark Matter candidate in the framework.
 
 \item Thus, we solve the Boltzmann equations that include the $N_{2,3}$ decays, inverse 
decays, and 2-to-2 scattering of matter in both sectors. These Boltzmann equations can be 
written in a schematic form as:
\begin{eqnarray}
\label{eq:BE}
\frac{s H_2}{z} Y_{N_{2,3}}' &=& -\gamma_{d_{2,3}} \left( 
\frac{Y_{N_{2,3}}}{Y_{N_{2,3}}^{\rm eq}} -1 \right) \, \,+ \, \, (2 
\leftrightarrow 2) \,, 
\\ \label{eq:BE2}
\frac{s H_2}{z} Y_{\Delta N_1}' &=& \gamma_{d_{2,3}} \left[ \epsilon_{D} \left( 
\frac{Y_{N_{2,3}}}{Y_{N_{2,3}}^{\rm eq}} -1 \right) - \frac{Y_{\Delta_ {D}}}{2 
Y_{D}^{\rm eq}} \, \mathrm{Br}_{D} \right]  \nonumber \\
&&\hspace*{0cm}+ \, \, (2 \leftrightarrow 2 ~\textrm{washout + transfer}) \,,
\label{eq:BE3}\\
\frac{s H_2}{z} Y_{\Delta l}' &=& \gamma_{d_{2,3}} \left[ \epsilon_l \left( 
\frac{Y_{N_{2,3}}}{Y_{N_{2,3}}^{\rm eq}} -1 \right) - \frac{Y_{\Delta l}}{2 
Y_l^{\rm eq}} \, \mbox{Br}_l\right] 
 \nonumber \\
&&\hspace*{0cm}+\, \, (2 \leftrightarrow 2 ~ \textrm{washout + transfer}) \,.
\label{eq:boltz4}
\end{eqnarray}
Here $z = M_{N_2}/T$,  $H_2$ is defined by \ref{eq:boltz6}, $s$ 
is the entropy density, $Y_{N_{2,3,l,D}}^{\rm eq}$ are the equilibrium number 
densities,
$\mathrm{Br}_{D/l}$ denote the branching fractions of $N_{2,3}$ into the two 
sectors and, $\gamma_{d_{2,3}}$ is the thermally averaged $N_{2,3}$ decay density and is 
expressed as:
\begin{equation}
 \gamma_{d_{2,3}}=\frac{M_{N_{2,3}}K_1(z)}{\pi^2z}\Gamma_{N_{2,3}}
\label{eq:boltz5}
\end{equation}

\item For simplicity, we focus only on the decay of $N_2$ to both sectors. 
The first term in Eq.(\ref{eq:BE3} , \ref{eq:boltz4}) is proportional to CP asymmetry ($\epsilon_l$,$\epsilon_{D}$), and the second term is proportional to Branching ratio $Br_{(x=l/D)}$ describing the effect of $2\rightarrow1$ inverse decay
processes leading to the washout of the asymmetries in each sector. The strength of the washout effects, as discussed in \cite{Falkowski:2011xh} is defined on the basis of $Br_x$ and $\Gamma_{N_2}$ values. Condition $Br_x \Gamma_{N_2}/H_2\ll 1$ marks the weak washout regime and 
$Br_x \Gamma_{N_2}/H_2\gg 1$ marks the strong washout regime. 
Here $H_2$ is the Hubble parameter at $T=M_{N_2}$, 
 \begin{equation}
  H_2=\sqrt{\frac{8\pi^3g*}{90}}.\frac{M_{N_2}^2}{\text{Mpl}}\simeq 
\mathcal{O}(10^4)
  \label{eq:boltz6}
 \end{equation}
with Mpl being the Planck mass in GeV. Detailed calculations of the relevant decay rates have been summarised in the appendix in a tabular form \ref{tab:decay}. From there, the branching ratio$(Br_{x}=\frac{\text{decay width into } 'x' \text{ sector}}{\text{total decay width}})$ for visible sector, $Br_L\simeq\mathcal{O}(10^0)$ with Decay width $\Gamma_{N_2}\simeq \mathcal{O}(10^5)$. Further, for Dark sector, $Br_D\simeq\mathcal{O}(10^{-2})$ with Decay width $\Gamma_{N_2}\simeq \mathcal{O}(10^{3})$. Based on the conditions for the washout regimes, data indicates that the dark sector lies in a weak washout regime, whereas the visible sector lies in a moderately-strong regime. For our case, we neglect the dynamics in the evolution of asymmetries due to 2-2 effects. Based on these conditions, Boltzmann equations~\ref{eq:BE}-\ref{eq:boltz4}, can be re-expressed as: 
\end{enumerate}  
\begin{eqnarray}
\frac{s H_2}{z} Y_{N_{2}}' &=& -\gamma_d \left( \frac{Y_{N_2}}{Y_{N_2}^{\rm eq}} 
-1 \right) \nonumber \\ 
\frac{s H_2}{z} Y_{\Delta N_1}' &=& \gamma_d \left[ \epsilon_{\rm D} \left( 
\frac{Y_{N_2}}{Y_{N_2}^{\rm eq}} -1 \nonumber \right)\right] \\
\frac{s H_2}{z} Y_{\Delta l}' &=& \gamma_d \left[ \epsilon_l \left( 
\frac{Y_{N_{2}}}{Y_{N_{2}}^{\rm eq}} -1 \right)- \frac{Y_{\Delta l}}{2 
Y_l^{\rm eq}} \, \mbox{Br}_l\right]
\label{eq:boltz7}
\end{eqnarray}
By putting all the numerical values into the Boltzmann Equations and solving it numerically 
using MATHEMATICA(13.1)~\cite{Mathematica}, the evolution of asymmetries in visible and dark sectors are obtained and can be referred from the plot~\ref{fig:dark2}. Also, using Eq.(\ref{eq:boltz1}), we plot a scatterplot~\ref{fig:scat3} depicting the allowed values of Dark Matter mass$(M_{N_1})$ and Dark sector asymmetry$(\epsilon_D)$ required to produce 
correct relic density of DM. 
 \section{Stability of Dark-matter}
 In this work, the lightest among three heavy RHN $N_1$ is the possible dark matter candidate which satisfies the exact relic abundance. In the original $\nu MSM$ model, $N_1$ has mass $\approx \mathcal{O}(keV)$ to $ \mathcal{O}(MeV)$. This DM can further decay to the light particles as follows:
 i)- via off-shell W/Z: $N_1\rightarrow l_1^-l_2^+\nu_{l2}$, $N_1\rightarrow l^-q_1\overline{q}_2$, $N_1\rightarrow l^-l^+\nu_{l}$, $N_1\rightarrow \nu_{l}\overline{l}'l'$, $N_1\rightarrow \nu_{l}q\overline{q}$, $N_1\rightarrow v_{l}v_{l}'\overline{v}_{l}'$, $N_1\rightarrow v_{l}v_{l}\overline{v}_{l}$ and ii)- via off-shell Higgs: $N_1\rightarrow v_{l}\overline{l}l$. The decay rates for these processes are given in table~\ref{tab:decay}.
 For $N_1$ to be a stable DM candidate, its lifetime should be more than the age of the universe. Out of all decays of $N_1$ the stronger constraint comes from $N_1 \rightarrow \gamma \nu$ \cite{Boyarsky:2009ix}. 
 \begin{equation}
  \label{eq:4}
   \Gamma_{N_1\to\gamma\nu} = \frac{9\, \alpha\, 
     G_F^2} {1024\pi^4}\sin^2(2\theta_1)\, M_1^5  
  \simeq 5.5\times10^{-22}\;\theta_1^2
  \left[\frac{M_1}{\mathrm{keV}}\right]^5\;\mathrm{s}^{-1}\;,
 \end{equation}
 where, $\alpha$ is fine structure constant, $G_F$ is Fermi's constant and $\theta_1^2=\Sigma_{i=1,2,3}\theta_{i1}^2$ is the active-sterile mixing angle.
 \begin{figure}
\centering
 \includegraphics[scale=0.6]{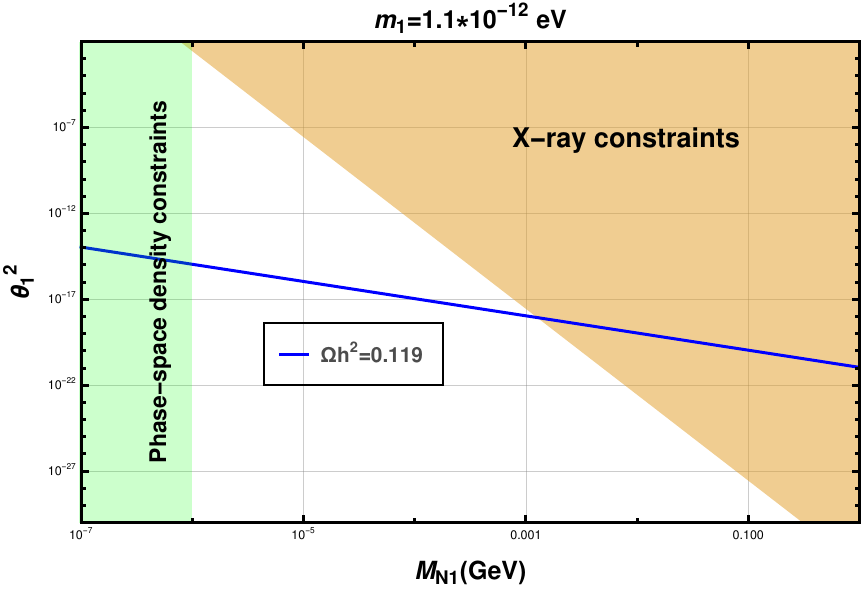}
 \caption{The figure shows constraints on $N_1$ mass and active-sterile mixing. 
 The yellow region (upper right corner) is excluded by X-ray observations
and the green region (rectangular block on the left), $M_1 < 1$keV by phase space analysis, which is the Tremaine-Gunn bound~\cite{Tremaine:1979we,Gorbunov:2008ka,Boyarsky:2008ju}. The points on the (blue) line, not excluded by other regions, give the correct $\Omega_{DM}$ in the present model.}
 \label{fig:xr}
\end{figure}
The allowed region of parameter space from X-ray observations in $\theta^2 - M_1$ plane can be referred from the plot in Fig:~\ref{fig:xr}. It can also be inferred from the Eq:~\ref{eq:4} that the entire allowed parameter space can accommodate a stable dark matter candidate. Thus, the range of DM masses allowed from plot~\ref{fig:scat3} satisfies both X-ray and phase space density constraints. Therefore, the DM candidate in our framework is stable without invoking any ad-hoc symmetry. 

\section{Detection Constraints}
\label{sec:det}
After putting in all the numerical values into the Boltzmann Equations and solving them, we show an asymmetric DM candidate $N_1$ being populated by the Cogenesis of asymmetry in both visible and dark sectors. The model allows for a DM mass range within the KeV scale. This mass range allows 
experimental and observational data from astrophysics, cosmology, and colliders to constrain the framework's parameter space. The constraints imposed on heavier mass DM($>$TeV) are already well-known and can be referred from~\cite{Jungman:1995df}. Also, it is established that if sterile neutrinos are produced via mixing with active neutrinos, then their mass is constrained within 0.4-50 KeV, irrespective of their production being resonant or non-resonant~\cite{Laine:2008pg,Boyarsky:2009ix}. Here, the lower bound seems to be fairly universal, but the upper bound depends more on the particular model framework setting to yield correct DM relic abundance and to also be consistent with the mixing angle constraints in relation to DM decays. In~\cite{Boyarsky:2012rt}, the various indirect detection techniques are discussed where the DM candidates are assumed to be majorly sterile neutrinos. It is argued that a dedicated cosmic mission- an X-ray spectrometer is needed to fully explore the signals of sterile neutrino-based DM. If such right-handed sterile neutrinos exist, they will affect the observed flux of neutrinos from a distant source, such as Sun or a supernova. Several experiments such as \cite{KATRIN:2018oow,Project8:2017nal,Blennow:2018hto,Super-Kamiokande:2014ndf,Wang:2015rma} are aimed to search for these fluxes. Some models predict that right-handed sterile neutrinos could mediate a process well-known in literature called neutrinoless double beta decay($\nu0\beta\beta$-decay). This process has not been observed yet, but several experiments, such as GERDA~\cite{GERDA:2013vls}, EXO~\cite{GERDA:2013vls}, and others~\cite{Klapdor-Kleingrothaus:2001oba} are searching for it, as their observation would directly indicate the presence of sterile neutrinos.\cite{Abada:2018qok,Dekens:2020ttz} and similar other works have studied the inter-relationship of KeV-scale sterile neutrinos and the $\nu0\beta\beta$-decay.

The mass of scalar singlet $\phi$ needs to be greater than the mass of $N_1$, so as to not provide a second DM candidate. Thus this puts a lower mass limit of $\sim \text{MeV}$ for $\phi$. The relic abundance of DM would contain the contribution from $\overline{N}_{1}$ DM component as well, but as the asymmetry in Dark Sector is of the order of $10^{-10}$, such a contribution is negligibly small and can be safely ignored. In here, as also provided in reference~\cite{Falkowski:2011xh}, there can exist interesting cosmological consequences depending on the mass range of $\phi$ singlet. Using~\cite{Jedamzik:2006xz,Kawasaki:2004qu}, we see that the only constraint required from BBN in a model with extra singlet scalar is the fast decay of $\phi$ into SM states to avoid late dissociation processes, but as there is no interaction of $\phi$ with SM particles in our framework, so such a constraint is evaded here. The presence of a singlet scalar means that any 
real or virtual Higgs production would frequently be associated with $\phi$ production, potentially leading to strong missing energy signals~\cite{Burgess:2000yq}. Any scalar portal mixing for a singlet scalar $\phi$ from a term like $\kappa_{\phi} (H^\dagger H)\phi$ in the lagrangian~\ref{eq:tf1} is suppressed by considering $\kappa_\phi$ to be negligibly small. As this is a scalar singlet extended SM framework, the constraints from astrophysics are generally less severe compared to a framework with SM extended with a $U(1)$ gauge group since a hidden photon couples to electric charge and, therefore equally to electrons and protons, but this can be avoided in an extended scalar framework. The nucleon-scalar$(\phi)$ elastic scattering cross section can also be obtained for a parameter space region and compared with the current limits for WIMP searches~\cite{Matos:2014gha}.
\par A totally sterile DM candidate can also be detected through its gravitational interactions with other matter. Several experiments, such as XENON and PandaX, are searching for this type of dark matter. For our model to be experimentally verifiable, data from experiments like~\cite{Wang:2021oha,CRESST:2020wtj,XENON100:2013ele}, could be analyzed to look for an excess flux of right-handed parity particles over their parity-inverted partners of BSM neutrinos($N_R$) in a specific comoving reference frame. Such a signal would hint towards a possible parity-asymmetry in RHNs. An analysis of the detection techniques for a FIMP DM can be found at \cite{Agrawal:2021dbo}. Also, the model incorporates the heavier RHNs at $10^{11-13}\text{ GeV}$ range, which via flavor leptogenesis scenario can explain the baryon asymmetry of the universe through sphaleron-like processes~\cite{Kuzmin:1985mm}. Right-handed sterile neutrinos could also interact very weakly with ordinary matter through weak interactions and neutrino mixings. Several laboratory experiments, such as MINOS~\cite{MINOS:2011amj} and LSND~\cite{LSND:2001aii}, target for evidence of such RHNs through these interactions. Articles \cite{Abazajian:2012ys,Deppisch:2015qwa,Dasgupta:2021ies,Beacham:2019nyx} sheds some light on the result of these experiments in the context of KeV scale sterile neutrinos. The model allows the CP violating decay of heavier RHNs into $N_1$ via scalar and vector bosons, which in turn are related to the entries of complex Yukawa matrix $y_{\nu}$ given in Eq.(\ref{eq:mnu_MR_def}). This gives a stringent upper limit on the allowed mass of the lightest active neutrino mass($m_1$) due to active-sterile mixing as given in reference~\cite{Datta:2021elq}. In summary, there are several proposed techniques for detecting right-handed sterile neutrinos, including oscillations, neutrinoless double beta decay, direct detection, and laboratory experiments. However, the existence of these particles has not been confirmed yet, though there have been hints towards excess signals at different mass scales.

\section{Conclusion}
\label{sec:con}
We have obtained the possibility of Asymmetric Dark Matter by an extension of the recently proposed model that accommodates FIMP within the $\nu$MSM paradigm. This has necessitated the introduction of a scalar $\phi$, but without invoking any new symmetries to stabilize the DM. The out-of-equilibrium decays of the heavier right-handed neutrinos are shown to generate both the matter anti-matter asymmetry of the visible sector through leptogenesis, as well as dark matter bearing a concomitant stamp of asymmetry. This is shown to be possible for
a wide range of DM masses, ranging from around 1 keV to a few hundred keVs consistent with the known light neutrino masses. This provides an exciting range of DM masses that can be probed for in conjunction with 
the signature of parity asymmetry in the Dark sector. This should be accessible to the direct and indirect search experiments that explore the ADM scenarios with the lightest right-handed neutrino $N_1$ as the Dark Matter. Other scenarios of leptogenesis, like resonant or Dirac leptogenesis, could also be tried within this setup to look for their relationship with the asymmetry in the two sectors. This, in turn, can have implications for the mass of the $N_1$ so as to be consistent with the visible sector asymmetry bounds.


\section*{Acknowledgement}
Utkarsh Patel(UP) would like to acknowledge the financial support obtained from the Ministry of Education, Government of India. Lekhika Malhotra(LM) would like to thank Dr. Sudhanwa Patra and the hospitality provided at IIT Bhilai during her stay there, which led to the completion of this project. UAY acknowledges an Institute Chair Professorship of IIT Bombay. We also thank Prof S. Umasankar for his valuable comments on the implications of our framework and Zafri Ahmed Borboruah for his helpful discussions.
\appendix
\label{app}
\section{Interaction Lagrangian in mass basis}
\label{app:il}
The flavor eigenstates and mass eigenstates are related to each other as 
follows,
\begin{eqnarray}
 &&\nu_{\alpha }= U_{\alpha i} \nu_{i}+S_{\alpha j} N^c_{j} \nonumber \\
 &&N_{\beta }=T^*_{\beta k} \nu^c_{k}+V^*_{\beta l}N_{l}
\end{eqnarray}
 
\subsection{Interactions of $N_R$ with Higgs}
\label{app:nh}
The relevant interaction of $N_R$ with SM Higgs:
 \begin{eqnarray}
-\mathcal{L}_{\rm Yuk}&\subset& \frac{g }{2 m_W} h \sum_{i,j=1}^{3} 
\bar{\nu}_{i}(U^{\dagger} S )_{ij}M_j N_{j}  \nonumber \\ &&
+ \frac{g }{2 m_W} h \sum_{i,j=1}^{3} \overline{N}^c_{i} (S^{\dagger}U)_{ij} 
m_j\nu^c_{j} \nonumber \\ && + \frac{g }{2 m_W} h \sum_{i,j=1}^{3}
 \overline{N}^c_{i}(S^{\dagger}S)_{ij}M_j N_{j} + h.c.,\label{ym} \nonumber
\end{eqnarray}

\subsection{Interactions of $N_R$ with $W,Z$}
\label{app:nw}
$N_R$ interaction with SM $W$ is understood in a flavor basis and is read as
\begin{eqnarray}
 \mathcal{L}^W_{CC}=\frac{g}{\sqrt{2}} (\overline{\nu_{\alpha 
L}}\slashed{W}^+l_{L\alpha} +\overline{l_{L\alpha}}\slashed{W}^-\nu_{L \alpha})
\end{eqnarray}
which can be reexpressed in a mass basis as follows,
\begin{eqnarray}
 \mathcal{L}^W_{CC} &=& - \frac{g}{\sqrt{2}} \sum_{\alpha,j=1}^{3} \overline{\ell}_j 
S_{ij}\gamma^{\mu} P_L N_j W_{\mu}^- + \mbox{h.c.} \quad \mbox{[mass 
basis]}
\end{eqnarray}

Similarly, the interaction of $N_R$ with neutral current interaction via SM $Z$ 
is given in flavor basis
\begin{eqnarray}
\mathcal{L}^Z_{NC} = \frac{g}{\cos \theta_W} \bigg(\frac{1}{2} 
\overline{\nu_{L\alpha}}\gamma^\mu \nu_{L\alpha}-\frac{1}{2}(c_w^2 - 
s_w^2)\overline{l_{L\alpha}}\gamma^\mu l_{L\alpha} \nonumber \\ 
+s_w^2\overline{l_{R\alpha}}\gamma^\mu l_{R\alpha} \bigg)Z_\mu \nonumber 
\end{eqnarray}
becomes, in mass basis, as
\begin{eqnarray}
-\mathcal{L_G}&\subset& \frac{g}{2 C_{\theta_w}} Z_{\mu} \sum_{i,j=1}^{3} 
\bar{\nu}_i (U^{\dagger}S)_{ij}\gamma^{\mu} P_L N^c_{j} \nonumber \\
&&+ \frac{g}{2 C_{\theta_w}} Z_{\mu} \sum_{i,j=1}^{3} \bar{N}^c_i 
(S^{\dagger}S)_{ij}\gamma^{\mu} P_L N^c_{j}  + h.c. +\cdots, \nonumber\\
\label{gauge_int}
\end{eqnarray}
\begin{table*}
\begin{center}
        \begin{tabular}{ | c |c | c | c | c |}
        \hline
            &Decays& $(x=1,y=0.1)$&$(x=0.8,y=0.08)$& $(x=0.6,y=0.001)$\\ \hline
           & $\Gamma_{N2\rightarrow l_\alpha W}$ & 0.016957 &0.013579&0.00982425\\
    Decay in Visible Sector     &$\Gamma_{N2\rightarrow \nu_\alpha Z}$& 
0.0171987&0.0137726&0.00996427\\
             &$\Gamma_{N2\rightarrow \nu_\alpha H}$& 0.016957&0.013579&0.00982425\\ 
\hline
              &$\Gamma_{N2\rightarrow N_1 H}$ & $1.09018\times 
10^{-26}$&$2.00607\times 10^{-26}$&$2.02565\times 10^{-26}$\\
               &$\Gamma_{N2\rightarrow N_1 Z}$& $8.42553\times 10^{-27}$ 
&$1.5504\times 10^{-26}$&$1.56553\times 10^{-26}$\\
&$\Gamma_{N2\rightarrow N_1 \phi}$&$0.000994718$&$0.000994718$&$0.000994718$\\ 
    Decay in Dark Sector  &$\Gamma_{W^+\rightarrow N_1 e^+}$ & $4.9719\times 
10^{-25}$&$4.9719\times 10^{-25}$&$4.9719\times 10^{-25}$\\
                  &$\Gamma_{Z\rightarrow N1 \overline{N}_1}$ & $7.53857\times 
10^{-49}$&$7.53857\times 10^{-49}$&$7.53857\times 10^{-49}$\\
                    &$\Gamma_{H\rightarrow N1 \overline{N}_1}$ &$6.95828\times 
10^{-54}$ &$6.95828\times 10^{-54}$&$6.95828\times 10^{-54}$\\ \hline
                      &$\Gamma_{N1\rightarrow l_\alpha^-l_\beta^+ \nu_\beta}(\alpha \neq \beta)$ 
&$1.71035\times 10^{-55}$&$1.71035\times 10^{-55}$&$1.71035\times 10^{-55}$\\
                        &$\Gamma_{N1\rightarrow  l_\alpha^- l_\alpha^+ \nu_\alpha}$ 
&$1.63207\times 10^{-55}$&$1.63207\times 10^{-55}$&$1.63207\times 10^{-55}$\\
     3body decay of $N_1$    &$\Gamma_{N1\rightarrow \nu_\alpha \overline{l_\beta} l_\beta 
}$ &$3.14933\times 10^{-56}$&$3.14933\times 10^{-56}$&$3.14933\times 
10^{-56}$\\
    &$\Gamma_{N1\rightarrow \nu_\alpha \nu_\beta 
\overline{\nu}_\beta }$ &$6.2842\times 10^{-56}$&$6.2842\times 
10^{-56}$&$6.2842\times 10^{-56}$\\
                           &$\Gamma_{N1\rightarrow q_{\alpha} q_{\alpha} \nu_\alpha 
}$ &$2.45034\times 10^{-55}$&$2.45034\times 10^{-55}$&$2.45034\times 10^{-55}$\\
        \hline
        \end{tabular}
       \end{center}
              \caption{Table shows the calculated numeric values for the decay rates of $N_2$ in visible as well as in the Dark sector and also the production and decay of $N_1$ in all the relevant scenarios for three different settings of angle variables x,y within the range given in plot~\ref{fig:scat1}. Readers can clearly see from the tabular data that the three body decays are negligible, making $N_1$ a stable dark matter candidate. Also, $\phi$-channel has a maximum contribution in the production of $N_1$ and hence creates the asymmetry channel in the Dark sector. }
        \label{tab:decay}
\end{table*}
\vspace{-0.3in}
\section{Decay Rates of RH-neutrinos $N_{k}$.}
\label{app:dr}
If mass of the heavy RH-neutrino $N_2$ greater than $W$, $Z$ and 
$H$, then using the interaction terms in mass basis, $N_2$ can decay into 
$H\nu$, $Z \nu$ and $W\ell$, respectively, if kinematically allowed. The two body decays are on-shell, although three body decays could be via off-shell SM bosons. The chosen values of important parameters are listed below:\\
\textbf{Input parameters:}
\begin{eqnarray}
   m_1 = 1.1\times10^{-21} ;\quad \Delta m_{atm} = 2.35*10^{-3}*10^{-18}; \Delta m_{sol} = 7.5*10^{-23} ;\nonumber \\ m_2=\sqrt{\Delta m_{sol}+m_1^2};m_3=\sqrt{\Delta m_{atm}+ m_1^2};\nonumber \\ M_1=10^{-5};\quad M_2=10^{11};\quad M_3=10^{12}; \quad \text{vsm}=246;\nonumber \\ \theta_{12} =33.48 {}^{\circ};\quad \theta_{23}=42.2 {}^{\circ};\quad \theta_{13} =8.52 {}^{\circ};\quad \delta =90 {}^{\circ};
 \end{eqnarray}where all masses are in GeV.
The expression for partial decay widths are shown below and their results are given in table~\ref{tab:decay}.
Also, the structure of coupling matrices are as follows:
\begin{equation}
y_{\nu} =\begin{pmatrix}
 3.5168083\times 10^{-16} & ~ 0.00159492 & ~ 0.00598712 \\
 1.7778870\times 10^{-16} & ~ 0.0062654 & ~ 0.00395143 \\
1.6273253\times 10^{-16} & ~ 0.00437137 & ~ 0.0168468 \\
\end{pmatrix}
\label{eq:Yukawa_coupling}
\end{equation}
and 
\begin{equation}
\lambda_{\phi N_i N_j} =\begin{pmatrix}
  10^{-3} & ~ 10^{-6} & ~ 10^{-8} \\
10^{-3} & ~ 10^{-6} & ~ 10^{-8} \\
 10^{-3} & ~10^{-6} & ~ 10^{-8} \\
\end{pmatrix}
\label{eq:Dark_coupling}
\end{equation}

\subsection*{Two body decay rates of $N_2$:}
\begin{enumerate}
\item  Decay of $N_2$ to visible sector-
\begin{equation}
\Gamma\big(N_{k} \to \ell_\alpha W \big) = 
\frac{g^2 \big|S_{\alpha k}\big|^2}{64 \pi} \,
\frac{\big(M^2_k -M^2_W \big)^2\,\big(M^2_k+ 2\,M^2_W \big)}{M^3_k\, M_W^2}
\end{equation}
\begin{equation}
 \Gamma\big(N_{k} \to \nu_\alpha Z \big) =\frac{g^2 \big|S_{\alpha k}\big|^2}{64 \pi \cos^2{\theta_W}}  \,
\frac{\big(M^2_k -M^2_Z \big)^2\,\big(M^2_k+ 2\,M^2_Z \big)}{M^3_k M_W^2}\, 
\end{equation}
\begin{equation}
\Gamma\big(N_{k} \to \nu_\alpha H \big) = \frac{g^2 \big|S_{\alpha 
k}\big|^2}{64 \pi} \,
\frac{\big(M^2_k -M^2_H \big)^2}{M_k\, M_W^2}
\end{equation}
\item Decay of $N_2$ to Dark sector-
 \begin{equation}
\Gamma(N_{k} \rightarrow N_1 H)= 
\frac{g^2 \big|S_{1 k}\big|^4}{64 \pi} \,
\frac{\big(M^2_k -M^2_H \big)^2}{M_k\, M_W^2} \,  
 \end{equation}
 \begin{equation}
\Gamma(N_{k} \rightarrow N_1 Z)= 
\frac{g^2 \big|S_{1 k}\big|^4}{64 \pi} \,
\frac{\big(M^2_k -M^2_Z \big)^2\,\big(M^2_k+ 2\,M^2_Z \big)}{M^3_k\, M_Z^2}  
 \end{equation}
  \begin{eqnarray}
  && \Gamma({N_k\rightarrow \phi N_{1}})=\Gamma1 = \frac{\lambda^2_{\phi N_1 N_1}}{32\pi M^3_{N_k}}\big( (M_{N_k} + m_{N_{1}})^2 - m^2_\phi \big)\times \nonumber \\ && \big( M^4_{N_k} + m^4_{N_{1}} + m^4_\phi - 2 M^2_{N_k} m^2_{N_{1}} - 2 m^2_{N_{1}} m^2_\phi - 2 M^2_{N_k} m^2_\phi \big)^{\frac{1}{2}}
 \end{eqnarray}
 \item Other processes that can produce dark matter $N_1$ are,
 \begin{equation} \Gamma(\phi\rightarrow N_1 \overline{N}_1)=\Gamma2= \frac{ m_{N_1}^2 m_{\phi} }{16 \pi v_{\text{SM}}^{ 2}} \left(1- \frac{4 m_{N_1}^2}{m_{\phi}^2}\right)^{3/2}
\label{eq:pdw1}
\end{equation}
The last two decay rates~\cite{Bandyopadhyay:2017bgh,Borah:2021pet} in \ref{eq:pdw1} correspond to the two vertices in figure~\ref{fig:dark2} going from left to right, respectively. The total decay rate will be,
\begin{equation}
\Gamma(N_2\rightarrow N_1N_1N_1)= \Gamma1+\Gamma2
\label{eq:pdw2}
\end{equation} 
\begin{equation}
 \Gamma(Z\rightarrow N_{1} \bar{N_1})= \frac{M_{Z}^3\,
\big|\lambda_{Z N_1 N_1} \big|^2}{24\pi\,v^2_{\rm 
SM}}\left(1-\frac{4M_{1}^2}{M_{Z}^2}\right)^{3/2}\,,
\label{zn1n1}
\end{equation}
\begin{equation}
 \Gamma(H \rightarrow N_1 \bar{N_1})= \frac{\lambda_{H N_1 N_1}^2\,M_H}
{16\pi}\left(1-\frac{4M_{1}^2}{M_{H}^2}\right)^{3/2}
\label{Hn1n1}\,
\end{equation}
\begin{multline}
\Gamma \big(W^+ \to N_1 e^+ \big) =  \frac{2\,(2M_{W}^{4}-(M_{e}^{2}-
M_{1}^{2})^{2}-(M_e^{2}+M_{1}^{2})M_{W}^{2})}{3v^2}\,\times\\ \big|\lambda_{W 
N_1 e} \big|^2\times\frac{\sqrt{1-\left( \frac{M_e+M_{1}}{M_{W}}\right)^{2}}
\sqrt{1-\left( \frac{M_e-M_{1}}{M_{W}}\right)^{2}}}{16\pi M_{W}} \,,
\label{wn1e} \nonumber
\end{multline}

\end{enumerate}

\subsection*{Three body decay rates of $N_2$:}
Interestingly, when RH-neutrinos are lighter than the SM bosons ($W,Z,H$), they 
eventually have three body decay modes with partial widths of $N_1$, in our present case is given by:
\begin{enumerate}
 \item For $N_{k} \to \ell^-_\alpha \ell^+_\beta \nu_{\beta}$ and $\alpha \neq \beta:$
 \begin{equation}
  \Gamma\big(N_{k} \to \ell^-_\alpha \ell^+_\alpha \nu_{\beta} 
\big) =  \Gamma\big(N_{k} \to \ell^+_\alpha \ell^-_\alpha \overline{\nu}_{\beta} 
\big) = \big|S_{\alpha k}\big|^2\,\frac{G^2_F}{192 \pi^3}\, 
M^5_{k}
 \end{equation}

 \item For $N_{k} \to \ell^-_\alpha \ell^+_\alpha \nu_{\alpha}:$
 \begin{equation}
\Gamma\big(N_{k} \to \ell^-_\alpha \ell^+_\alpha \nu_{\alpha} 
\big) =  \Gamma\big(N_{k} \to \ell^-_\alpha \ell^+_\alpha \overline{\nu}_{\alpha} 
\big) = \big|S_{\alpha k}\big|^2\,\frac{G^2_F}{192 \pi^3}\,
M^5_{k} \big( (1+C_L^2) +C_R^2 \big)
 \end{equation}
 \item For $N_1 \to \nu _{\alpha } \overline{l}_{\beta } l_{\beta }$
 \begin{equation}
   \Gamma\big(N_1 \to \nu _{\alpha } \overline{l}_{\beta } l_{\beta } 
\big) =  \Gamma\big(N_1 \to \overline{\nu} _{\alpha } l_{\beta } \overline{l}_{\beta }
\big) = \big|S_{\alpha k}\big|^2\,\frac{G^2_F}{192 \pi^3}\, 
M^5_{k} (C_L^2+C_R^2)
 \end{equation}
\item For, $N_1 \to \nu _{\alpha } \nu _{\beta } \overline{\nu} _{\beta }:$
 \begin{eqnarray}
 &&   \Gamma\big(N_1 \to \nu _{\alpha } \nu _{\beta } \overline{\nu} _{\beta}
 \big) =  \Gamma\big(N_1 \to \overline{\nu} _{\alpha } \overline{\nu} _{\beta } \nu _{\beta } 
\big) = \big| S_{\alpha k}\big|^2 \, \frac{G^2_F}{192 \pi^3}\, 
 M^5_{k} C_{\nu}^2 \quad \mbox{\text{for} $\alpha \neq \beta$} \\
 && \hspace*{3.9cm}  = \big| S_{\alpha k}\big|^2 \, \frac{G^2_F}{192 \pi^3}\, 
 M^5_{k} 4 C_{\nu}^2  \quad \mbox{\text{for} $\alpha = \beta$}
 \end{eqnarray}
 \item For $N_k\to \nu _{\alpha } q_{\alpha } q_{\alpha }:$ 
 \begin{equation}
  \Gamma\big(N_k\to \nu _{\alpha } q_{\alpha } \overline{q}_{\alpha }
\big) =  \Gamma\big(N_k\to \overline{\nu} _{\alpha } \overline{q}_{\alpha } q_{\alpha }
\big) = \big|S_{\alpha k}\big|^2\,\frac{G^2_F}{192 \pi^3}\, 
M^5_{k} N_C [(C^q_L)^2+(C^q_R)^2]
 \end{equation}
  \item For, $N_k \to \ell_\alpha^-  q_a \bar q_b :$
  \begin{equation}
 \Gamma \big(N_k \to \ell_\alpha^-  q_a \bar q_b\big)  = 
\Gamma\big(N_k \to \ell_\alpha^+  \bar q_a  q_b\big) 
 \approx  N_C \big|S_{\alpha k}\big|^2 
\big|V_{CKM}^{ab}\big|^2 \frac{G_F^2}{192 \pi^3} M_{k}^5,
  \end{equation}
  where, $N_C = 3$ denotes the number of color degrees of freedom of quarks. $C_{\nu}=\frac{1}{2}$ and 
  \begin{eqnarray}
    C_L =-\frac{1}{2}+ \sin^2\theta_W ,\quad C_R= \sin^2\theta_W, \nonumber\\
   C_L^u=\frac{1}{2} -\frac{2}{3} \sin^2 \theta_W, \, \, C_R^u= -\frac{2}{3} \sin^2 \theta_W,
\nonumber \\
C_L^d = -\frac{1}{2}+\frac{1}{3} \sin^2 \theta_W, \, \, C_R^d= \frac{1}{3}\sin^2 \theta_W, 
  \end{eqnarray}
\end{enumerate}

\bibliography{adm.bib}

\begin{thebibliography}{83}%
\makeatletter
\providecommand \@ifxundefined [1]{%
 \@ifx{#1\undefined}
}%
\providecommand \@ifnum [1]{%
 \ifnum #1\expandafter \@firstoftwo
 \else \expandafter \@secondoftwo
 \fi
}%
\providecommand \@ifx [1]{%
 \ifx #1\expandafter \@firstoftwo
 \else \expandafter \@secondoftwo
 \fi
}%
\providecommand \natexlab [1]{#1}%
\providecommand \enquote  [1]{``#1''}%
\providecommand \bibnamefont  [1]{#1}%
\providecommand \bibfnamefont [1]{#1}%
\providecommand \citenamefont [1]{#1}%
\providecommand \href@noop [0]{\@secondoftwo}%
\providecommand \href [0]{\begingroup \@sanitize@url \@href}%
\providecommand \@href[1]{\@@startlink{#1}\@@href}%
\providecommand \@@href[1]{\endgroup#1\@@endlink}%
\providecommand \@sanitize@url [0]{\catcode `\\12\catcode `\$12\catcode
  `\&12\catcode `\#12\catcode `\^12\catcode `\_12\catcode `\%12\relax}%
\providecommand \@@startlink[1]{}%
\providecommand \@@endlink[0]{}%
\providecommand \url  [0]{\begingroup\@sanitize@url \@url }%
\providecommand \@url [1]{\endgroup\@href {#1}{\urlprefix }}%
\providecommand \urlprefix  [0]{URL }%
\providecommand \Eprint [0]{\href }%
\providecommand \doibase [0]{http://dx.doi.org/}%
\providecommand \selectlanguage [0]{\@gobble}%
\providecommand \bibinfo  [0]{\@secondoftwo}%
\providecommand \bibfield  [0]{\@secondoftwo}%
\providecommand \translation [1]{[#1]}%
\providecommand \BibitemOpen [0]{}%
\providecommand \bibitemStop [0]{}%
\providecommand \bibitemNoStop [0]{.\EOS\space}%
\providecommand \EOS [0]{\spacefactor3000\relax}%
\providecommand \BibitemShut  [1]{\csname bibitem#1\endcsname}%
\let\auto@bib@innerbib\@empty
\bibitem [{\citenamefont {Datta}\ \emph {et~al.}(2021)\citenamefont {Datta},
  \citenamefont {Roshan},\ and\ \citenamefont {Sil}}]{Datta:2021elq}%
  \BibitemOpen
  \bibfield  {author} {\bibinfo {author} {\bibfnamefont {A.}~\bibnamefont
  {Datta}}, \bibinfo {author} {\bibfnamefont {R.}~\bibnamefont {Roshan}}, \
  and\ \bibinfo {author} {\bibfnamefont {A.}~\bibnamefont {Sil}},\ }\href
  {\doibase 10.1103/PhysRevLett.127.231801} {\bibfield  {journal} {\bibinfo
  {journal} {Phys. Rev. Lett.}\ }\textbf {\bibinfo {volume} {127}},\ \bibinfo
  {pages} {231801} (\bibinfo {year} {2021})},\ \Eprint
  {http://arxiv.org/abs/2104.02030} {arXiv:2104.02030 [hep-ph]} \BibitemShut
  {NoStop}%
\bibitem [{\citenamefont {Fukuda}\ \emph {et~al.}(1998)\citenamefont {Fukuda}
  \emph {et~al.}}]{Super-Kamiokande:1998kpq}%
  \BibitemOpen
  \bibfield  {author} {\bibinfo {author} {\bibfnamefont {Y.}~\bibnamefont
  {Fukuda}} \emph {et~al.} (\bibinfo {collaboration} {Super-Kamiokande}),\
  }\href {\doibase 10.1103/PhysRevLett.81.1562} {\bibfield  {journal} {\bibinfo
   {journal} {Phys. Rev. Lett.}\ }\textbf {\bibinfo {volume} {81}},\ \bibinfo
  {pages} {1562} (\bibinfo {year} {1998})},\ \Eprint
  {http://arxiv.org/abs/hep-ex/9807003} {arXiv:hep-ex/9807003} \BibitemShut
  {NoStop}%
\bibitem [{\citenamefont {Ahmad}\ \emph {et~al.}(2002)\citenamefont {Ahmad}
  \emph {et~al.}}]{SNO:2002tuh}%
  \BibitemOpen
  \bibfield  {author} {\bibinfo {author} {\bibfnamefont {Q.~R.}\ \bibnamefont
  {Ahmad}} \emph {et~al.} (\bibinfo {collaboration} {SNO}),\ }\href {\doibase
  10.1103/PhysRevLett.89.011301} {\bibfield  {journal} {\bibinfo  {journal}
  {Phys. Rev. Lett.}\ }\textbf {\bibinfo {volume} {89}},\ \bibinfo {pages}
  {011301} (\bibinfo {year} {2002})},\ \Eprint
  {http://arxiv.org/abs/nucl-ex/0204008} {arXiv:nucl-ex/0204008} \BibitemShut
  {NoStop}%
\bibitem [{\citenamefont {Ahn}\ \emph {et~al.}(2003)\citenamefont {Ahn} \emph
  {et~al.}}]{K2K:2002icj}%
  \BibitemOpen
  \bibfield  {author} {\bibinfo {author} {\bibfnamefont {M.~H.}\ \bibnamefont
  {Ahn}} \emph {et~al.} (\bibinfo {collaboration} {K2K}),\ }\href {\doibase
  10.1103/PhysRevLett.90.041801} {\bibfield  {journal} {\bibinfo  {journal}
  {Phys. Rev. Lett.}\ }\textbf {\bibinfo {volume} {90}},\ \bibinfo {pages}
  {041801} (\bibinfo {year} {2003})},\ \Eprint
  {http://arxiv.org/abs/hep-ex/0212007} {arXiv:hep-ex/0212007} \BibitemShut
  {NoStop}%
\bibitem [{\citenamefont {Julian}(1967)}]{Julian:1967zz}%
  \BibitemOpen
  \bibfield  {author} {\bibinfo {author} {\bibfnamefont {W.~H.}\ \bibnamefont
  {Julian}},\ }\href {\doibase 10.1086/149134} {\bibfield  {journal} {\bibinfo
  {journal} {Astrophys. J.}\ }\textbf {\bibinfo {volume} {148}},\ \bibinfo
  {pages} {175} (\bibinfo {year} {1967})}\BibitemShut {NoStop}%
\bibitem [{\citenamefont {Tegmark}\ \emph {et~al.}(2004)\citenamefont {Tegmark}
  \emph {et~al.}}]{SDSS:2003eyi}%
  \BibitemOpen
  \bibfield  {author} {\bibinfo {author} {\bibfnamefont {M.}~\bibnamefont
  {Tegmark}} \emph {et~al.} (\bibinfo {collaboration} {SDSS}),\ }\href
  {\doibase 10.1103/PhysRevD.69.103501} {\bibfield  {journal} {\bibinfo
  {journal} {Phys. Rev. D}\ }\textbf {\bibinfo {volume} {69}},\ \bibinfo
  {pages} {103501} (\bibinfo {year} {2004})},\ \Eprint
  {http://arxiv.org/abs/astro-ph/0310723} {arXiv:astro-ph/0310723} \BibitemShut
  {NoStop}%
\bibitem [{\citenamefont {Aghanim}\ \emph {et~al.}(2020)\citenamefont {Aghanim}
  \emph {et~al.}}]{Planck:2018vyg}%
  \BibitemOpen
  \bibfield  {author} {\bibinfo {author} {\bibfnamefont {N.}~\bibnamefont
  {Aghanim}} \emph {et~al.} (\bibinfo {collaboration} {Planck}),\ }\href
  {\doibase 10.1051/0004-6361/201833910} {\bibfield  {journal} {\bibinfo
  {journal} {Astron. Astrophys.}\ }\textbf {\bibinfo {volume} {641}},\ \bibinfo
  {pages} {A6} (\bibinfo {year} {2020})},\ \bibinfo {note} {[Erratum:
  Astron.Astrophys. 652, C4 (2021)]},\ \Eprint
  {http://arxiv.org/abs/1807.06209} {arXiv:1807.06209 [astro-ph.CO]}
  \BibitemShut {NoStop}%
\bibitem [{\citenamefont {Asaka}\ and\ \citenamefont
  {Shaposhnikov}(2005)}]{Asaka:2005pn}%
  \BibitemOpen
  \bibfield  {author} {\bibinfo {author} {\bibfnamefont {T.}~\bibnamefont
  {Asaka}}\ and\ \bibinfo {author} {\bibfnamefont {M.}~\bibnamefont
  {Shaposhnikov}},\ }\href {\doibase 10.1016/j.physletb.2005.06.020} {\bibfield
   {journal} {\bibinfo  {journal} {Phys. Lett. B}\ }\textbf {\bibinfo {volume}
  {620}},\ \bibinfo {pages} {17} (\bibinfo {year} {2005})},\ \Eprint
  {http://arxiv.org/abs/hep-ph/0505013} {arXiv:hep-ph/0505013} \BibitemShut
  {NoStop}%
\bibitem [{\citenamefont {Asaka}\ \emph {et~al.}(2005)\citenamefont {Asaka},
  \citenamefont {Blanchet},\ and\ \citenamefont {Shaposhnikov}}]{Asaka:2005an}%
  \BibitemOpen
  \bibfield  {author} {\bibinfo {author} {\bibfnamefont {T.}~\bibnamefont
  {Asaka}}, \bibinfo {author} {\bibfnamefont {S.}~\bibnamefont {Blanchet}}, \
  and\ \bibinfo {author} {\bibfnamefont {M.}~\bibnamefont {Shaposhnikov}},\
  }\href {\doibase 10.1016/j.physletb.2005.09.070} {\bibfield  {journal}
  {\bibinfo  {journal} {Phys. Lett. B}\ }\textbf {\bibinfo {volume} {631}},\
  \bibinfo {pages} {151} (\bibinfo {year} {2005})},\ \Eprint
  {http://arxiv.org/abs/hep-ph/0503065} {arXiv:hep-ph/0503065} \BibitemShut
  {NoStop}%
\bibitem [{\citenamefont {Sahu}\ and\ \citenamefont
  {Yajnik}(2006)}]{Sahu:2005fe}%
  \BibitemOpen
  \bibfield  {author} {\bibinfo {author} {\bibfnamefont {N.}~\bibnamefont
  {Sahu}}\ and\ \bibinfo {author} {\bibfnamefont {U.~A.}\ \bibnamefont
  {Yajnik}},\ }\href {\doibase 10.1016/j.physletb.2006.02.040} {\bibfield
  {journal} {\bibinfo  {journal} {Phys. Lett. B}\ }\textbf {\bibinfo {volume}
  {635}},\ \bibinfo {pages} {11} (\bibinfo {year} {2006})},\ \Eprint
  {http://arxiv.org/abs/hep-ph/0509285} {arXiv:hep-ph/0509285} \BibitemShut
  {NoStop}%
\bibitem [{\citenamefont {Canetti}\ \emph {et~al.}(2013)\citenamefont
  {Canetti}, \citenamefont {Drewes}, \citenamefont {Frossard},\ and\
  \citenamefont {Shaposhnikov}}]{Canetti:2012kh}%
  \BibitemOpen
  \bibfield  {author} {\bibinfo {author} {\bibfnamefont {L.}~\bibnamefont
  {Canetti}}, \bibinfo {author} {\bibfnamefont {M.}~\bibnamefont {Drewes}},
  \bibinfo {author} {\bibfnamefont {T.}~\bibnamefont {Frossard}}, \ and\
  \bibinfo {author} {\bibfnamefont {M.}~\bibnamefont {Shaposhnikov}},\ }\href
  {\doibase 10.1103/PhysRevD.87.093006} {\bibfield  {journal} {\bibinfo
  {journal} {Phys. Rev. D}\ }\textbf {\bibinfo {volume} {87}},\ \bibinfo
  {pages} {093006} (\bibinfo {year} {2013})},\ \Eprint
  {http://arxiv.org/abs/1208.4607} {arXiv:1208.4607 [hep-ph]} \BibitemShut
  {NoStop}%
\bibitem [{\citenamefont {Yoshimura}(1978)}]{Yoshimura:1978ex}%
  \BibitemOpen
  \bibfield  {author} {\bibinfo {author} {\bibfnamefont {M.}~\bibnamefont
  {Yoshimura}},\ }\href {\doibase 10.1103/PhysRevLett.41.281} {\bibfield
  {journal} {\bibinfo  {journal} {Phys. Rev. Lett.}\ }\textbf {\bibinfo
  {volume} {41}},\ \bibinfo {pages} {281} (\bibinfo {year} {1978})},\ \bibinfo
  {note} {[Erratum: Phys.Rev.Lett. 42, 746 (1979)]}\BibitemShut {NoStop}%
\bibitem [{\citenamefont {Weinberg}(1979)}]{Weinberg:1979bt}%
  \BibitemOpen
  \bibfield  {author} {\bibinfo {author} {\bibfnamefont {S.}~\bibnamefont
  {Weinberg}},\ }\href {\doibase 10.1103/PhysRevLett.42.850} {\bibfield
  {journal} {\bibinfo  {journal} {Phys. Rev. Lett.}\ }\textbf {\bibinfo
  {volume} {42}},\ \bibinfo {pages} {850} (\bibinfo {year} {1979})}\BibitemShut
  {NoStop}%
\bibitem [{\citenamefont {Fukugita}\ and\ \citenamefont
  {Yanagida}(1986)}]{Fukugita:1986hr}%
  \BibitemOpen
  \bibfield  {author} {\bibinfo {author} {\bibfnamefont {M.}~\bibnamefont
  {Fukugita}}\ and\ \bibinfo {author} {\bibfnamefont {T.}~\bibnamefont
  {Yanagida}},\ }\href {\doibase 10.1016/0370-2693(86)91126-3} {\bibfield
  {journal} {\bibinfo  {journal} {Phys. Lett. B}\ }\textbf {\bibinfo {volume}
  {174}},\ \bibinfo {pages} {45} (\bibinfo {year} {1986})}\BibitemShut
  {NoStop}%
\bibitem [{\citenamefont {Luty}(1992)}]{Luty:1992un}%
  \BibitemOpen
  \bibfield  {author} {\bibinfo {author} {\bibfnamefont {M.~A.}\ \bibnamefont
  {Luty}},\ }\href {\doibase 10.1103/PhysRevD.45.455} {\bibfield  {journal}
  {\bibinfo  {journal} {Phys. Rev. D}\ }\textbf {\bibinfo {volume} {45}},\
  \bibinfo {pages} {455} (\bibinfo {year} {1992})}\BibitemShut {NoStop}%
\bibitem [{\citenamefont {Kuzmin}\ \emph {et~al.}(1985)\citenamefont {Kuzmin},
  \citenamefont {Rubakov},\ and\ \citenamefont {Shaposhnikov}}]{Kuzmin:1985mm}%
  \BibitemOpen
  \bibfield  {author} {\bibinfo {author} {\bibfnamefont {V.~A.}\ \bibnamefont
  {Kuzmin}}, \bibinfo {author} {\bibfnamefont {V.~A.}\ \bibnamefont {Rubakov}},
  \ and\ \bibinfo {author} {\bibfnamefont {M.~E.}\ \bibnamefont
  {Shaposhnikov}},\ }\href@noop {} {\bibfield  {journal} {\bibinfo  {journal}
  {Phys. Lett.}\ }\textbf {\bibinfo {volume} {B155}},\ \bibinfo {pages} {36}
  (\bibinfo {year} {1985})}\BibitemShut {NoStop}%
\bibitem [{\citenamefont {Spergel}\ \emph {et~al.}(2003)\citenamefont {Spergel}
  \emph {et~al.}}]{WMAP:2003elm}%
  \BibitemOpen
  \bibfield  {author} {\bibinfo {author} {\bibfnamefont {D.~N.}\ \bibnamefont
  {Spergel}} \emph {et~al.} (\bibinfo {collaboration} {WMAP}),\ }\href
  {\doibase 10.1086/377226} {\bibfield  {journal} {\bibinfo  {journal}
  {Astrophys. J. Suppl.}\ }\textbf {\bibinfo {volume} {148}},\ \bibinfo {pages}
  {175} (\bibinfo {year} {2003})},\ \Eprint
  {http://arxiv.org/abs/astro-ph/0302209} {arXiv:astro-ph/0302209} \BibitemShut
  {NoStop}%
\bibitem [{\citenamefont {Ade}\ \emph {et~al.}(2016)\citenamefont {Ade} \emph
  {et~al.}}]{Planck:2015fie}%
  \BibitemOpen
  \bibfield  {author} {\bibinfo {author} {\bibfnamefont {P.~A.~R.}\
  \bibnamefont {Ade}} \emph {et~al.} (\bibinfo {collaboration} {Planck}),\
  }\href {\doibase 10.1051/0004-6361/201525830} {\bibfield  {journal} {\bibinfo
   {journal} {Astron. Astrophys.}\ }\textbf {\bibinfo {volume} {594}},\
  \bibinfo {pages} {A13} (\bibinfo {year} {2016})},\ \Eprint
  {http://arxiv.org/abs/1502.01589} {arXiv:1502.01589 [astro-ph.CO]}
  \BibitemShut {NoStop}%
\bibitem [{\citenamefont {Nussinov}(1985)}]{Nussinov:1985xr}%
  \BibitemOpen
  \bibfield  {author} {\bibinfo {author} {\bibfnamefont {S.}~\bibnamefont
  {Nussinov}},\ }\href {\doibase 10.1016/0370-2693(85)90689-6} {\bibfield
  {journal} {\bibinfo  {journal} {Phys. Lett. B}\ }\textbf {\bibinfo {volume}
  {165}},\ \bibinfo {pages} {55} (\bibinfo {year} {1985})}\BibitemShut
  {NoStop}%
\bibitem [{\citenamefont {Kaplan}(1992)}]{Kaplan:1991ah}%
  \BibitemOpen
  \bibfield  {author} {\bibinfo {author} {\bibfnamefont {D.~B.}\ \bibnamefont
  {Kaplan}},\ }\href {\doibase 10.1103/PhysRevLett.68.741} {\bibfield
  {journal} {\bibinfo  {journal} {Phys. Rev. Lett.}\ }\textbf {\bibinfo
  {volume} {68}},\ \bibinfo {pages} {741} (\bibinfo {year} {1992})}\BibitemShut
  {NoStop}%
\bibitem [{\citenamefont {Foot}(2014)}]{Foot:2013msa}%
  \BibitemOpen
  \bibfield  {author} {\bibinfo {author} {\bibfnamefont {R.}~\bibnamefont
  {Foot}},\ }\href {\doibase 10.1016/j.physletb.2013.11.019} {\bibfield
  {journal} {\bibinfo  {journal} {Phys. Lett. B}\ }\textbf {\bibinfo {volume}
  {728}},\ \bibinfo {pages} {45} (\bibinfo {year} {2014})},\ \Eprint
  {http://arxiv.org/abs/1305.4316} {arXiv:1305.4316 [astro-ph.CO]} \BibitemShut
  {NoStop}%
\bibitem [{\citenamefont {Blinnikov}\ and\ \citenamefont
  {Khlopov}(1982)}]{Blinnikov:1982eh}%
  \BibitemOpen
  \bibfield  {author} {\bibinfo {author} {\bibfnamefont {S.~I.}\ \bibnamefont
  {Blinnikov}}\ and\ \bibinfo {author} {\bibfnamefont {M.~Y.}\ \bibnamefont
  {Khlopov}},\ }\href@noop {} {\bibfield  {journal} {\bibinfo  {journal} {Sov.
  J. Nucl. Phys.}\ }\textbf {\bibinfo {volume} {36}},\ \bibinfo {pages} {472}
  (\bibinfo {year} {1982})}\BibitemShut {NoStop}%
\bibitem [{\citenamefont {Berezhiani}\ \emph {et~al.}(2005)\citenamefont
  {Berezhiani}, \citenamefont {Ciarcelluti}, \citenamefont {Comelli},\ and\
  \citenamefont {Villante}}]{Berezhiani:2003wj}%
  \BibitemOpen
  \bibfield  {author} {\bibinfo {author} {\bibfnamefont {Z.}~\bibnamefont
  {Berezhiani}}, \bibinfo {author} {\bibfnamefont {P.}~\bibnamefont
  {Ciarcelluti}}, \bibinfo {author} {\bibfnamefont {D.}~\bibnamefont
  {Comelli}}, \ and\ \bibinfo {author} {\bibfnamefont {F.~L.}\ \bibnamefont
  {Villante}},\ }\href {\doibase 10.1142/S0218271805005165} {\bibfield
  {journal} {\bibinfo  {journal} {Int. J. Mod. Phys. D}\ }\textbf {\bibinfo
  {volume} {14}},\ \bibinfo {pages} {107} (\bibinfo {year} {2005})},\ \Eprint
  {http://arxiv.org/abs/astro-ph/0312605} {arXiv:astro-ph/0312605} \BibitemShut
  {NoStop}%
\bibitem [{\citenamefont {Ciarcelluti}(2005)}]{Ciarcelluti:2004ip}%
  \BibitemOpen
  \bibfield  {author} {\bibinfo {author} {\bibfnamefont {P.}~\bibnamefont
  {Ciarcelluti}},\ }\href {\doibase 10.1142/S0218271805006225} {\bibfield
  {journal} {\bibinfo  {journal} {Int. J. Mod. Phys. D}\ }\textbf {\bibinfo
  {volume} {14}},\ \bibinfo {pages} {223} (\bibinfo {year} {2005})},\ \Eprint
  {http://arxiv.org/abs/astro-ph/0409633} {arXiv:astro-ph/0409633} \BibitemShut
  {NoStop}%
\bibitem [{\citenamefont {Hooper}\ \emph {et~al.}(2005)\citenamefont {Hooper},
  \citenamefont {March-Russell},\ and\ \citenamefont {West}}]{Hooper:2004dc}%
  \BibitemOpen
  \bibfield  {author} {\bibinfo {author} {\bibfnamefont {D.}~\bibnamefont
  {Hooper}}, \bibinfo {author} {\bibfnamefont {J.}~\bibnamefont
  {March-Russell}}, \ and\ \bibinfo {author} {\bibfnamefont {S.~M.}\
  \bibnamefont {West}},\ }\href {\doibase 10.1016/j.physletb.2004.11.047}
  {\bibfield  {journal} {\bibinfo  {journal} {Phys. Lett. B}\ }\textbf
  {\bibinfo {volume} {605}},\ \bibinfo {pages} {228} (\bibinfo {year}
  {2005})},\ \Eprint {http://arxiv.org/abs/hep-ph/0410114}
  {arXiv:hep-ph/0410114} \BibitemShut {NoStop}%
\bibitem [{\citenamefont {Kaplan}\ \emph {et~al.}(2009)\citenamefont {Kaplan},
  \citenamefont {Luty},\ and\ \citenamefont {Zurek}}]{Kaplan:2009ag}%
  \BibitemOpen
  \bibfield  {author} {\bibinfo {author} {\bibfnamefont {D.~E.}\ \bibnamefont
  {Kaplan}}, \bibinfo {author} {\bibfnamefont {M.~A.}\ \bibnamefont {Luty}}, \
  and\ \bibinfo {author} {\bibfnamefont {K.~M.}\ \bibnamefont {Zurek}},\ }\href
  {\doibase 10.1103/PhysRevD.79.115016} {\bibfield  {journal} {\bibinfo
  {journal} {Phys. Rev. D}\ }\textbf {\bibinfo {volume} {79}},\ \bibinfo
  {pages} {115016} (\bibinfo {year} {2009})},\ \Eprint
  {http://arxiv.org/abs/0901.4117} {arXiv:0901.4117 [hep-ph]} \BibitemShut
  {NoStop}%
\bibitem [{\citenamefont {Haba}\ \emph {et~al.}(2011)\citenamefont {Haba},
  \citenamefont {Matsumoto},\ and\ \citenamefont {Sato}}]{Haba:2011uz}%
  \BibitemOpen
  \bibfield  {author} {\bibinfo {author} {\bibfnamefont {N.}~\bibnamefont
  {Haba}}, \bibinfo {author} {\bibfnamefont {S.}~\bibnamefont {Matsumoto}}, \
  and\ \bibinfo {author} {\bibfnamefont {R.}~\bibnamefont {Sato}},\ }\href
  {\doibase 10.1103/PhysRevD.84.055016} {\bibfield  {journal} {\bibinfo
  {journal} {Phys. Rev. D}\ }\textbf {\bibinfo {volume} {84}},\ \bibinfo
  {pages} {055016} (\bibinfo {year} {2011})},\ \Eprint
  {http://arxiv.org/abs/1101.5679} {arXiv:1101.5679 [hep-ph]} \BibitemShut
  {NoStop}%
\bibitem [{\citenamefont {Dutta}\ and\ \citenamefont
  {Kumar}(2006)}]{Dutta:2006pt}%
  \BibitemOpen
  \bibfield  {author} {\bibinfo {author} {\bibfnamefont {B.}~\bibnamefont
  {Dutta}}\ and\ \bibinfo {author} {\bibfnamefont {J.}~\bibnamefont {Kumar}},\
  }\href {\doibase 10.1016/j.physletb.2006.09.069} {\bibfield  {journal}
  {\bibinfo  {journal} {Phys. Lett. B}\ }\textbf {\bibinfo {volume} {643}},\
  \bibinfo {pages} {284} (\bibinfo {year} {2006})},\ \Eprint
  {http://arxiv.org/abs/hep-th/0608188} {arXiv:hep-th/0608188} \BibitemShut
  {NoStop}%
\bibitem [{\citenamefont {Shelton}\ and\ \citenamefont
  {Zurek}(2010)}]{Shelton:2010ta}%
  \BibitemOpen
  \bibfield  {author} {\bibinfo {author} {\bibfnamefont {J.}~\bibnamefont
  {Shelton}}\ and\ \bibinfo {author} {\bibfnamefont {K.~M.}\ \bibnamefont
  {Zurek}},\ }\href {\doibase 10.1103/PhysRevD.82.123512} {\bibfield  {journal}
  {\bibinfo  {journal} {Phys. Rev. D}\ }\textbf {\bibinfo {volume} {82}},\
  \bibinfo {pages} {123512} (\bibinfo {year} {2010})},\ \Eprint
  {http://arxiv.org/abs/1008.1997} {arXiv:1008.1997 [hep-ph]} \BibitemShut
  {NoStop}%
\bibitem [{\citenamefont {Falkowski}\ \emph {et~al.}(2011)\citenamefont
  {Falkowski}, \citenamefont {Ruderman},\ and\ \citenamefont
  {Volansky}}]{Falkowski:2011xh}%
  \BibitemOpen
  \bibfield  {author} {\bibinfo {author} {\bibfnamefont {A.}~\bibnamefont
  {Falkowski}}, \bibinfo {author} {\bibfnamefont {J.~T.}\ \bibnamefont
  {Ruderman}}, \ and\ \bibinfo {author} {\bibfnamefont {T.}~\bibnamefont
  {Volansky}},\ }\href {\doibase 10.1007/JHEP05(2011)106} {\bibfield  {journal}
  {\bibinfo  {journal} {JHEP}\ }\textbf {\bibinfo {volume} {05}},\ \bibinfo
  {pages} {106} (\bibinfo {year} {2011})},\ \Eprint
  {http://arxiv.org/abs/1101.4936} {arXiv:1101.4936 [hep-ph]} \BibitemShut
  {NoStop}%
\bibitem [{\citenamefont {von Harling}\ \emph {et~al.}(2012)\citenamefont {von
  Harling}, \citenamefont {Petraki},\ and\ \citenamefont
  {Volkas}}]{vonHarling:2012yn}%
  \BibitemOpen
  \bibfield  {author} {\bibinfo {author} {\bibfnamefont {B.}~\bibnamefont {von
  Harling}}, \bibinfo {author} {\bibfnamefont {K.}~\bibnamefont {Petraki}}, \
  and\ \bibinfo {author} {\bibfnamefont {R.~R.}\ \bibnamefont {Volkas}},\
  }\href {\doibase 10.1088/1475-7516/2012/05/021} {\bibfield  {journal}
  {\bibinfo  {journal} {JCAP}\ }\textbf {\bibinfo {volume} {05}},\ \bibinfo
  {pages} {021} (\bibinfo {year} {2012})},\ \Eprint
  {http://arxiv.org/abs/1201.2200} {arXiv:1201.2200 [hep-ph]} \BibitemShut
  {NoStop}%
\bibitem [{\citenamefont {Bell}\ \emph {et~al.}(2011)\citenamefont {Bell},
  \citenamefont {Petraki}, \citenamefont {Shoemaker},\ and\ \citenamefont
  {Volkas}}]{Bell:2011tn}%
  \BibitemOpen
  \bibfield  {author} {\bibinfo {author} {\bibfnamefont {N.~F.}\ \bibnamefont
  {Bell}}, \bibinfo {author} {\bibfnamefont {K.}~\bibnamefont {Petraki}},
  \bibinfo {author} {\bibfnamefont {I.~M.}\ \bibnamefont {Shoemaker}}, \ and\
  \bibinfo {author} {\bibfnamefont {R.~R.}\ \bibnamefont {Volkas}},\ }\href
  {\doibase 10.1103/PhysRevD.84.123505} {\bibfield  {journal} {\bibinfo
  {journal} {Phys. Rev. D}\ }\textbf {\bibinfo {volume} {84}},\ \bibinfo
  {pages} {123505} (\bibinfo {year} {2011})},\ \Eprint
  {http://arxiv.org/abs/1105.3730} {arXiv:1105.3730 [hep-ph]} \BibitemShut
  {NoStop}%
\bibitem [{\citenamefont {Kuzmin}(1998)}]{Kuzmin:1996he}%
  \BibitemOpen
  \bibfield  {author} {\bibinfo {author} {\bibfnamefont {V.~A.}\ \bibnamefont
  {Kuzmin}},\ }\href {\doibase 10.1134/1.953070} {\bibfield  {journal}
  {\bibinfo  {journal} {Phys. Part. Nucl.}\ }\textbf {\bibinfo {volume} {29}},\
  \bibinfo {pages} {257} (\bibinfo {year} {1998})},\ \Eprint
  {http://arxiv.org/abs/hep-ph/9701269} {arXiv:hep-ph/9701269} \BibitemShut
  {NoStop}%
\bibitem [{\citenamefont {Dulaney}\ \emph {et~al.}(2011)\citenamefont
  {Dulaney}, \citenamefont {Fileviez~Perez},\ and\ \citenamefont
  {Wise}}]{Dulaney:2010dj}%
  \BibitemOpen
  \bibfield  {author} {\bibinfo {author} {\bibfnamefont {T.~R.}\ \bibnamefont
  {Dulaney}}, \bibinfo {author} {\bibfnamefont {P.}~\bibnamefont
  {Fileviez~Perez}}, \ and\ \bibinfo {author} {\bibfnamefont {M.~B.}\
  \bibnamefont {Wise}},\ }\href {\doibase 10.1103/PhysRevD.83.023520}
  {\bibfield  {journal} {\bibinfo  {journal} {Phys. Rev. D}\ }\textbf {\bibinfo
  {volume} {83}},\ \bibinfo {pages} {023520} (\bibinfo {year} {2011})},\
  \Eprint {http://arxiv.org/abs/1005.0617} {arXiv:1005.0617 [hep-ph]}
  \BibitemShut {NoStop}%
\bibitem [{\citenamefont {An}\ \emph {et~al.}(2010)\citenamefont {An},
  \citenamefont {Chen}, \citenamefont {Mohapatra},\ and\ \citenamefont
  {Zhang}}]{An:2009vq}%
  \BibitemOpen
  \bibfield  {author} {\bibinfo {author} {\bibfnamefont {H.}~\bibnamefont
  {An}}, \bibinfo {author} {\bibfnamefont {S.-L.}\ \bibnamefont {Chen}},
  \bibinfo {author} {\bibfnamefont {R.~N.}\ \bibnamefont {Mohapatra}}, \ and\
  \bibinfo {author} {\bibfnamefont {Y.}~\bibnamefont {Zhang}},\ }\href
  {\doibase 10.1007/JHEP03(2010)124} {\bibfield  {journal} {\bibinfo  {journal}
  {JHEP}\ }\textbf {\bibinfo {volume} {03}},\ \bibinfo {pages} {124} (\bibinfo
  {year} {2010})},\ \Eprint {http://arxiv.org/abs/0911.4463} {arXiv:0911.4463
  [hep-ph]} \BibitemShut {NoStop}%
\bibitem [{\citenamefont {Farrar}\ and\ \citenamefont
  {Zaharijas}(2006)}]{Farrar:2005zd}%
  \BibitemOpen
  \bibfield  {author} {\bibinfo {author} {\bibfnamefont {G.~R.}\ \bibnamefont
  {Farrar}}\ and\ \bibinfo {author} {\bibfnamefont {G.}~\bibnamefont
  {Zaharijas}},\ }\href {\doibase 10.1103/PhysRevLett.96.041302} {\bibfield
  {journal} {\bibinfo  {journal} {Phys. Rev. Lett.}\ }\textbf {\bibinfo
  {volume} {96}},\ \bibinfo {pages} {041302} (\bibinfo {year} {2006})},\
  \Eprint {http://arxiv.org/abs/hep-ph/0510079} {arXiv:hep-ph/0510079}
  \BibitemShut {NoStop}%
\bibitem [{\citenamefont {Bhattacharya}\ \emph {et~al.}(2021)\citenamefont
  {Bhattacharya}, \citenamefont {Roshan}, \citenamefont {Sil},\ and\
  \citenamefont {Vatsyayan}}]{Bhattacharya:2021jli}%
  \BibitemOpen
  \bibfield  {author} {\bibinfo {author} {\bibfnamefont {S.}~\bibnamefont
  {Bhattacharya}}, \bibinfo {author} {\bibfnamefont {R.}~\bibnamefont
  {Roshan}}, \bibinfo {author} {\bibfnamefont {A.}~\bibnamefont {Sil}}, \ and\
  \bibinfo {author} {\bibfnamefont {D.}~\bibnamefont {Vatsyayan}},\ }\href@noop
  {} {\  (\bibinfo {year} {2021})},\ \Eprint {http://arxiv.org/abs/2105.06189}
  {arXiv:2105.06189 [hep-ph]} \BibitemShut {NoStop}%
\bibitem [{\citenamefont {Minkowski}(1977)}]{Minkowski:1977sc}%
  \BibitemOpen
  \bibfield  {author} {\bibinfo {author} {\bibfnamefont {P.}~\bibnamefont
  {Minkowski}},\ }\href {\doibase 10.1016/0370-2693(77)90435-X} {\bibfield
  {journal} {\bibinfo  {journal} {Phys. Lett. B}\ }\textbf {\bibinfo {volume}
  {67}},\ \bibinfo {pages} {421} (\bibinfo {year} {1977})}\BibitemShut
  {NoStop}%
\bibitem [{\citenamefont {Mohapatra}\ and\ \citenamefont
  {Senjanovic}(1980)}]{Mohapatra:1979ia}%
  \BibitemOpen
  \bibfield  {author} {\bibinfo {author} {\bibfnamefont {R.~N.}\ \bibnamefont
  {Mohapatra}}\ and\ \bibinfo {author} {\bibfnamefont {G.}~\bibnamefont
  {Senjanovic}},\ }\href {\doibase 10.1103/PhysRevLett.44.912} {\bibfield
  {journal} {\bibinfo  {journal} {Phys. Rev. Lett.}\ }\textbf {\bibinfo
  {volume} {44}},\ \bibinfo {pages} {912} (\bibinfo {year} {1980})}\BibitemShut
  {NoStop}%
\bibitem [{\citenamefont {Yanagida}(1979)}]{Yanagida:1979as}%
  \BibitemOpen
  \bibfield  {author} {\bibinfo {author} {\bibfnamefont {T.}~\bibnamefont
  {Yanagida}},\ }\href@noop {} {\bibfield  {journal} {\bibinfo  {journal}
  {Conf. Proc. C}\ }\textbf {\bibinfo {volume} {7902131}},\ \bibinfo {pages}
  {95} (\bibinfo {year} {1979})}\BibitemShut {NoStop}%
\bibitem [{\citenamefont {Schechter}\ and\ \citenamefont
  {Valle}(1980)}]{Schechter:1980gr}%
  \BibitemOpen
  \bibfield  {author} {\bibinfo {author} {\bibfnamefont {J.}~\bibnamefont
  {Schechter}}\ and\ \bibinfo {author} {\bibfnamefont {J.~W.~F.}\ \bibnamefont
  {Valle}},\ }\href {\doibase 10.1103/PhysRevD.22.2227} {\bibfield  {journal}
  {\bibinfo  {journal} {Phys. Rev. D}\ }\textbf {\bibinfo {volume} {22}},\
  \bibinfo {pages} {2227} (\bibinfo {year} {1980})}\BibitemShut {NoStop}%
\bibitem [{\citenamefont {Branco}\ \emph {et~al.}(2020)\citenamefont {Branco},
  \citenamefont {Penedo}, \citenamefont {Pereira}, \citenamefont {Rebelo},\
  and\ \citenamefont {Silva-Marcos}}]{Branco:2020yvs}%
  \BibitemOpen
  \bibfield  {author} {\bibinfo {author} {\bibfnamefont {G.~C.}\ \bibnamefont
  {Branco}}, \bibinfo {author} {\bibfnamefont {J.~T.}\ \bibnamefont {Penedo}},
  \bibinfo {author} {\bibfnamefont {P.~M.~F.}\ \bibnamefont {Pereira}},
  \bibinfo {author} {\bibfnamefont {M.~N.}\ \bibnamefont {Rebelo}}, \ and\
  \bibinfo {author} {\bibfnamefont {J.~I.}\ \bibnamefont {Silva-Marcos}},\
  }\href {\doibase 10.1007/JHEP07(2020)164} {\bibfield  {journal} {\bibinfo
  {journal} {JHEP}\ }\textbf {\bibinfo {volume} {07}},\ \bibinfo {pages} {164}
  (\bibinfo {year} {2020})},\ \Eprint {http://arxiv.org/abs/1912.05875}
  {arXiv:1912.05875 [hep-ph]} \BibitemShut {NoStop}%
\bibitem [{\citenamefont {Sakharov}(1967)}]{Sakharov:1967dj}%
  \BibitemOpen
  \bibfield  {author} {\bibinfo {author} {\bibfnamefont {A.~D.}\ \bibnamefont
  {Sakharov}},\ }\href {\doibase 10.1070/PU1991v034n05ABEH002497} {\bibfield
  {journal} {\bibinfo  {journal} {Pisma Zh. Eksp. Teor. Fiz.}\ }\textbf
  {\bibinfo {volume} {5}},\ \bibinfo {pages} {32} (\bibinfo {year}
  {1967})}\BibitemShut {NoStop}%
\bibitem [{\citenamefont {Affleck}\ and\ \citenamefont
  {Dine}(1985)}]{Affleck:1984fy}%
  \BibitemOpen
  \bibfield  {author} {\bibinfo {author} {\bibfnamefont {I.}~\bibnamefont
  {Affleck}}\ and\ \bibinfo {author} {\bibfnamefont {M.}~\bibnamefont {Dine}},\
  }\href {\doibase 10.1016/0550-3213(85)90021-5} {\bibfield  {journal}
  {\bibinfo  {journal} {Nucl. Phys. B}\ }\textbf {\bibinfo {volume} {249}},\
  \bibinfo {pages} {361} (\bibinfo {year} {1985})}\BibitemShut {NoStop}%
\bibitem [{\citenamefont {Davidson}\ \emph {et~al.}(2008)\citenamefont
  {Davidson}, \citenamefont {Nardi},\ and\ \citenamefont
  {Nir}}]{Davidson:2008bu}%
  \BibitemOpen
  \bibfield  {author} {\bibinfo {author} {\bibfnamefont {S.}~\bibnamefont
  {Davidson}}, \bibinfo {author} {\bibfnamefont {E.}~\bibnamefont {Nardi}}, \
  and\ \bibinfo {author} {\bibfnamefont {Y.}~\bibnamefont {Nir}},\ }\href
  {\doibase 10.1016/j.physrep.2008.06.002} {\bibfield  {journal} {\bibinfo
  {journal} {Phys. Rept.}\ }\textbf {\bibinfo {volume} {466}},\ \bibinfo
  {pages} {105} (\bibinfo {year} {2008})},\ \Eprint
  {http://arxiv.org/abs/0802.2962} {arXiv:0802.2962 [hep-ph]} \BibitemShut
  {NoStop}%
\bibitem [{\citenamefont {Buchmuller}\ \emph {et~al.}(2004)\citenamefont
  {Buchmuller}, \citenamefont {Di~Bari},\ and\ \citenamefont
  {Plumacher}}]{Buchmuller:2004tu}%
  \BibitemOpen
  \bibfield  {author} {\bibinfo {author} {\bibfnamefont {W.}~\bibnamefont
  {Buchmuller}}, \bibinfo {author} {\bibfnamefont {P.}~\bibnamefont {Di~Bari}},
  \ and\ \bibinfo {author} {\bibfnamefont {M.}~\bibnamefont {Plumacher}},\
  }\href {\doibase 10.1088/1367-2630/6/1/105} {\bibfield  {journal} {\bibinfo
  {journal} {New J. Phys.}\ }\textbf {\bibinfo {volume} {6}},\ \bibinfo {pages}
  {105} (\bibinfo {year} {2004})},\ \Eprint
  {http://arxiv.org/abs/hep-ph/0406014} {arXiv:hep-ph/0406014} \BibitemShut
  {NoStop}%
\bibitem [{\citenamefont {Blanchet}(2008)}]{Blanchet:2008hg}%
  \BibitemOpen
  \bibfield  {author} {\bibinfo {author} {\bibfnamefont {S.}~\bibnamefont
  {Blanchet}},\ }\emph {\bibinfo {title} {{A New Era of Leptogenesis}}},\
  \href@noop {} {\bibinfo {type} {Other thesis}} (\bibinfo {year} {2008}),\
  \Eprint {http://arxiv.org/abs/0807.1408} {arXiv:0807.1408 [hep-ph]}
  \BibitemShut {NoStop}%
\bibitem [{\citenamefont {Hagedorn}\ \emph {et~al.}(2018)\citenamefont
  {Hagedorn}, \citenamefont {Mohapatra}, \citenamefont {Molinaro},
  \citenamefont {Nishi},\ and\ \citenamefont {Petcov}}]{Hagedorn:2017wjy}%
  \BibitemOpen
  \bibfield  {author} {\bibinfo {author} {\bibfnamefont {C.}~\bibnamefont
  {Hagedorn}}, \bibinfo {author} {\bibfnamefont {R.~N.}\ \bibnamefont
  {Mohapatra}}, \bibinfo {author} {\bibfnamefont {E.}~\bibnamefont {Molinaro}},
  \bibinfo {author} {\bibfnamefont {C.~C.}\ \bibnamefont {Nishi}}, \ and\
  \bibinfo {author} {\bibfnamefont {S.~T.}\ \bibnamefont {Petcov}},\ }\href
  {\doibase 10.1142/S0217751X1842006X} {\bibfield  {journal} {\bibinfo
  {journal} {Int. J. Mod. Phys. A}\ }\textbf {\bibinfo {volume} {33}},\
  \bibinfo {pages} {1842006} (\bibinfo {year} {2018})},\ \Eprint
  {http://arxiv.org/abs/1711.02866} {arXiv:1711.02866 [hep-ph]} \BibitemShut
  {NoStop}%
\bibitem [{\citenamefont {Iminniyaz}\ \emph {et~al.}(2011)\citenamefont
  {Iminniyaz}, \citenamefont {Drees},\ and\ \citenamefont
  {Chen}}]{Iminniyaz:2011yp}%
  \BibitemOpen
  \bibfield  {author} {\bibinfo {author} {\bibfnamefont {H.}~\bibnamefont
  {Iminniyaz}}, \bibinfo {author} {\bibfnamefont {M.}~\bibnamefont {Drees}}, \
  and\ \bibinfo {author} {\bibfnamefont {X.}~\bibnamefont {Chen}},\ }\href
  {\doibase 10.1088/1475-7516/2011/07/003} {\bibfield  {journal} {\bibinfo
  {journal} {JCAP}\ }\textbf {\bibinfo {volume} {07}},\ \bibinfo {pages} {003}
  (\bibinfo {year} {2011})},\ \Eprint {http://arxiv.org/abs/1104.5548}
  {arXiv:1104.5548 [hep-ph]} \BibitemShut {NoStop}%
\bibitem [{\citenamefont {Kolb}\ and\ \citenamefont
  {Turner}(1990)}]{Kolb:1990vq}%
  \BibitemOpen
  \bibfield  {author} {\bibinfo {author} {\bibfnamefont {E.~W.}\ \bibnamefont
  {Kolb}}\ and\ \bibinfo {author} {\bibfnamefont {M.~S.}\ \bibnamefont
  {Turner}},\ }\href {\doibase 10.1201/9780429492860} {\emph {\bibinfo {title}
  {{The Early Universe}}}},\ Vol.~\bibinfo {volume} {69}\ (\bibinfo {year}
  {1990})\BibitemShut {NoStop}%
\bibitem [{\citenamefont {Inc.}()}]{Mathematica}%
  \BibitemOpen
  \bibfield  {author} {\bibinfo {author} {\bibfnamefont {W.~R.}\ \bibnamefont
  {Inc.}},\ }\href {https://www.wolfram.com/mathematica} {\enquote {\bibinfo
  {title} {Mathematica, {V}ersion 13.1},}\ }\bibinfo {note} {Champaign, IL,
  2022}\BibitemShut {NoStop}%
\bibitem [{\citenamefont {Boyarsky}\ \emph
  {et~al.}(2009{\natexlab{a}})\citenamefont {Boyarsky}, \citenamefont
  {Ruchayskiy},\ and\ \citenamefont {Shaposhnikov}}]{Boyarsky:2009ix}%
  \BibitemOpen
  \bibfield  {author} {\bibinfo {author} {\bibfnamefont {A.}~\bibnamefont
  {Boyarsky}}, \bibinfo {author} {\bibfnamefont {O.}~\bibnamefont
  {Ruchayskiy}}, \ and\ \bibinfo {author} {\bibfnamefont {M.}~\bibnamefont
  {Shaposhnikov}},\ }\href {\doibase 10.1146/annurev.nucl.010909.083654}
  {\bibfield  {journal} {\bibinfo  {journal} {Ann. Rev. Nucl. Part. Sci.}\
  }\textbf {\bibinfo {volume} {59}},\ \bibinfo {pages} {191} (\bibinfo {year}
  {2009}{\natexlab{a}})},\ \Eprint {http://arxiv.org/abs/0901.0011}
  {arXiv:0901.0011 [hep-ph]} \BibitemShut {NoStop}%
\bibitem [{\citenamefont {Tremaine}\ and\ \citenamefont
  {Gunn}(1979)}]{Tremaine:1979we}%
  \BibitemOpen
  \bibfield  {author} {\bibinfo {author} {\bibfnamefont {S.}~\bibnamefont
  {Tremaine}}\ and\ \bibinfo {author} {\bibfnamefont {J.~E.}\ \bibnamefont
  {Gunn}},\ }\href {\doibase 10.1103/PhysRevLett.42.407} {\bibfield  {journal}
  {\bibinfo  {journal} {Phys. Rev. Lett.}\ }\textbf {\bibinfo {volume} {42}},\
  \bibinfo {pages} {407} (\bibinfo {year} {1979})}\BibitemShut {NoStop}%
\bibitem [{\citenamefont {Gorbunov}\ \emph {et~al.}(2008)\citenamefont
  {Gorbunov}, \citenamefont {Khmelnitsky},\ and\ \citenamefont
  {Rubakov}}]{Gorbunov:2008ka}%
  \BibitemOpen
  \bibfield  {author} {\bibinfo {author} {\bibfnamefont {D.}~\bibnamefont
  {Gorbunov}}, \bibinfo {author} {\bibfnamefont {A.}~\bibnamefont
  {Khmelnitsky}}, \ and\ \bibinfo {author} {\bibfnamefont {V.}~\bibnamefont
  {Rubakov}},\ }\href {\doibase 10.1088/1475-7516/2008/10/041} {\bibfield
  {journal} {\bibinfo  {journal} {JCAP}\ }\textbf {\bibinfo {volume} {10}},\
  \bibinfo {pages} {041} (\bibinfo {year} {2008})},\ \Eprint
  {http://arxiv.org/abs/0808.3910} {arXiv:0808.3910 [hep-ph]} \BibitemShut
  {NoStop}%
\bibitem [{\citenamefont {Boyarsky}\ \emph
  {et~al.}(2009{\natexlab{b}})\citenamefont {Boyarsky}, \citenamefont
  {Ruchayskiy},\ and\ \citenamefont {Iakubovskyi}}]{Boyarsky:2008ju}%
  \BibitemOpen
  \bibfield  {author} {\bibinfo {author} {\bibfnamefont {A.}~\bibnamefont
  {Boyarsky}}, \bibinfo {author} {\bibfnamefont {O.}~\bibnamefont
  {Ruchayskiy}}, \ and\ \bibinfo {author} {\bibfnamefont {D.}~\bibnamefont
  {Iakubovskyi}},\ }\href {\doibase 10.1088/1475-7516/2009/03/005} {\bibfield
  {journal} {\bibinfo  {journal} {JCAP}\ }\textbf {\bibinfo {volume} {03}},\
  \bibinfo {pages} {005} (\bibinfo {year} {2009}{\natexlab{b}})},\ \Eprint
  {http://arxiv.org/abs/0808.3902} {arXiv:0808.3902 [hep-ph]} \BibitemShut
  {NoStop}%
\bibitem [{\citenamefont {Jungman}\ \emph {et~al.}(1996)\citenamefont
  {Jungman}, \citenamefont {Kamionkowski},\ and\ \citenamefont
  {Griest}}]{Jungman:1995df}%
  \BibitemOpen
  \bibfield  {author} {\bibinfo {author} {\bibfnamefont {G.}~\bibnamefont
  {Jungman}}, \bibinfo {author} {\bibfnamefont {M.}~\bibnamefont
  {Kamionkowski}}, \ and\ \bibinfo {author} {\bibfnamefont {K.}~\bibnamefont
  {Griest}},\ }\href {\doibase 10.1016/0370-1573(95)00058-5} {\bibfield
  {journal} {\bibinfo  {journal} {Phys. Rept.}\ }\textbf {\bibinfo {volume}
  {267}},\ \bibinfo {pages} {195} (\bibinfo {year} {1996})},\ \Eprint
  {http://arxiv.org/abs/hep-ph/9506380} {arXiv:hep-ph/9506380} \BibitemShut
  {NoStop}%
\bibitem [{\citenamefont {Laine}\ and\ \citenamefont
  {Shaposhnikov}(2008)}]{Laine:2008pg}%
  \BibitemOpen
  \bibfield  {author} {\bibinfo {author} {\bibfnamefont {M.}~\bibnamefont
  {Laine}}\ and\ \bibinfo {author} {\bibfnamefont {M.}~\bibnamefont
  {Shaposhnikov}},\ }\href {\doibase 10.1088/1475-7516/2008/06/031} {\bibfield
  {journal} {\bibinfo  {journal} {JCAP}\ }\textbf {\bibinfo {volume} {06}},\
  \bibinfo {pages} {031} (\bibinfo {year} {2008})},\ \Eprint
  {http://arxiv.org/abs/0804.4543} {arXiv:0804.4543 [hep-ph]} \BibitemShut
  {NoStop}%
\bibitem [{\citenamefont {Boyarsky}\ \emph {et~al.}(2012)\citenamefont
  {Boyarsky}, \citenamefont {Iakubovskyi},\ and\ \citenamefont
  {Ruchayskiy}}]{Boyarsky:2012rt}%
  \BibitemOpen
  \bibfield  {author} {\bibinfo {author} {\bibfnamefont {A.}~\bibnamefont
  {Boyarsky}}, \bibinfo {author} {\bibfnamefont {D.}~\bibnamefont
  {Iakubovskyi}}, \ and\ \bibinfo {author} {\bibfnamefont {O.}~\bibnamefont
  {Ruchayskiy}},\ }\href {\doibase 10.1016/j.dark.2012.11.001} {\bibfield
  {journal} {\bibinfo  {journal} {Phys. Dark Univ.}\ }\textbf {\bibinfo
  {volume} {1}},\ \bibinfo {pages} {136} (\bibinfo {year} {2012})},\ \Eprint
  {http://arxiv.org/abs/1306.4954} {arXiv:1306.4954 [astro-ph.CO]} \BibitemShut
  {NoStop}%
\bibitem [{\citenamefont {Mertens}\ \emph {et~al.}(2019)\citenamefont {Mertens}
  \emph {et~al.}}]{KATRIN:2018oow}%
  \BibitemOpen
  \bibfield  {author} {\bibinfo {author} {\bibfnamefont {S.}~\bibnamefont
  {Mertens}} \emph {et~al.} (\bibinfo {collaboration} {KATRIN}),\ }\href
  {\doibase 10.1088/1361-6471/ab12fe} {\bibfield  {journal} {\bibinfo
  {journal} {J. Phys. G}\ }\textbf {\bibinfo {volume} {46}},\ \bibinfo {pages}
  {065203} (\bibinfo {year} {2019})},\ \Eprint
  {http://arxiv.org/abs/1810.06711} {arXiv:1810.06711 [physics.ins-det]}
  \BibitemShut {NoStop}%
\bibitem [{\citenamefont {Ashtari~Esfahani}\ \emph {et~al.}(2017)\citenamefont
  {Ashtari~Esfahani} \emph {et~al.}}]{Project8:2017nal}%
  \BibitemOpen
  \bibfield  {author} {\bibinfo {author} {\bibfnamefont {A.}~\bibnamefont
  {Ashtari~Esfahani}} \emph {et~al.} (\bibinfo {collaboration} {Project 8}),\
  }\href {\doibase 10.1088/1361-6471/aa5b4f} {\bibfield  {journal} {\bibinfo
  {journal} {J. Phys. G}\ }\textbf {\bibinfo {volume} {44}},\ \bibinfo {pages}
  {054004} (\bibinfo {year} {2017})},\ \Eprint
  {http://arxiv.org/abs/1703.02037} {arXiv:1703.02037 [physics.ins-det]}
  \BibitemShut {NoStop}%
\bibitem [{\citenamefont {Blennow}\ \emph {et~al.}(2018)\citenamefont
  {Blennow}, \citenamefont {Fernandez-Martinez}, \citenamefont {Gehrlein},
  \citenamefont {Hernandez-Garcia},\ and\ \citenamefont
  {Salvado}}]{Blennow:2018hto}%
  \BibitemOpen
  \bibfield  {author} {\bibinfo {author} {\bibfnamefont {M.}~\bibnamefont
  {Blennow}}, \bibinfo {author} {\bibfnamefont {E.}~\bibnamefont
  {Fernandez-Martinez}}, \bibinfo {author} {\bibfnamefont {J.}~\bibnamefont
  {Gehrlein}}, \bibinfo {author} {\bibfnamefont {J.}~\bibnamefont
  {Hernandez-Garcia}}, \ and\ \bibinfo {author} {\bibfnamefont
  {J.}~\bibnamefont {Salvado}},\ }\href {\doibase
  10.1140/epjc/s10052-018-6282-2} {\bibfield  {journal} {\bibinfo  {journal}
  {Eur. Phys. J. C}\ }\textbf {\bibinfo {volume} {78}},\ \bibinfo {pages} {807}
  (\bibinfo {year} {2018})},\ \Eprint {http://arxiv.org/abs/1803.02362}
  {arXiv:1803.02362 [hep-ph]} \BibitemShut {NoStop}%
\bibitem [{\citenamefont {Abe}\ \emph {et~al.}(2015)\citenamefont {Abe} \emph
  {et~al.}}]{Super-Kamiokande:2014ndf}%
  \BibitemOpen
  \bibfield  {author} {\bibinfo {author} {\bibfnamefont {K.}~\bibnamefont
  {Abe}} \emph {et~al.} (\bibinfo {collaboration} {Super-Kamiokande}),\ }\href
  {\doibase 10.1103/PhysRevD.91.052019} {\bibfield  {journal} {\bibinfo
  {journal} {Phys. Rev. D}\ }\textbf {\bibinfo {volume} {91}},\ \bibinfo
  {pages} {052019} (\bibinfo {year} {2015})},\ \Eprint
  {http://arxiv.org/abs/1410.2008} {arXiv:1410.2008 [hep-ex]} \BibitemShut
  {NoStop}%
\bibitem [{\citenamefont {Wang}\ and\ \citenamefont {Xing}()}]{Wang:2015rma}%
  \BibitemOpen
  \bibfield  {author} {\bibinfo {author} {\bibfnamefont {Y.}~\bibnamefont
  {Wang}}\ and\ \bibinfo {author} {\bibfnamefont {Z.-z.}\ \bibnamefont
  {Xing}},\ }\href@noop {} {\ }\Eprint {http://arxiv.org/abs/1504.06155}
  {arXiv:1504.06155 [hep-ph]} \BibitemShut {NoStop}%
\bibitem [{\citenamefont {Agostini}\ \emph {et~al.}(2013)\citenamefont
  {Agostini} \emph {et~al.}}]{GERDA:2013vls}%
  \BibitemOpen
  \bibfield  {author} {\bibinfo {author} {\bibfnamefont {M.}~\bibnamefont
  {Agostini}} \emph {et~al.} (\bibinfo {collaboration} {GERDA}),\ }\href
  {\doibase 10.1103/PhysRevLett.111.122503} {\bibfield  {journal} {\bibinfo
  {journal} {Phys. Rev. Lett.}\ }\textbf {\bibinfo {volume} {111}},\ \bibinfo
  {pages} {122503} (\bibinfo {year} {2013})},\ \Eprint
  {http://arxiv.org/abs/1307.4720} {arXiv:1307.4720 [nucl-ex]} \BibitemShut
  {NoStop}%
\bibitem [{\citenamefont {Klapdor-Kleingrothaus}\ \emph
  {et~al.}(2001)\citenamefont {Klapdor-Kleingrothaus}, \citenamefont {Dietz},
  \citenamefont {Harney},\ and\ \citenamefont
  {Krivosheina}}]{Klapdor-Kleingrothaus:2001oba}%
  \BibitemOpen
  \bibfield  {author} {\bibinfo {author} {\bibfnamefont {H.~V.}\ \bibnamefont
  {Klapdor-Kleingrothaus}}, \bibinfo {author} {\bibfnamefont {A.}~\bibnamefont
  {Dietz}}, \bibinfo {author} {\bibfnamefont {H.~L.}\ \bibnamefont {Harney}}, \
  and\ \bibinfo {author} {\bibfnamefont {I.~V.}\ \bibnamefont {Krivosheina}},\
  }\href {\doibase 10.1142/S0217732301005825} {\bibfield  {journal} {\bibinfo
  {journal} {Mod. Phys. Lett. A}\ }\textbf {\bibinfo {volume} {16}},\ \bibinfo
  {pages} {2409} (\bibinfo {year} {2001})},\ \Eprint
  {http://arxiv.org/abs/hep-ph/0201231} {arXiv:hep-ph/0201231} \BibitemShut
  {NoStop}%
\bibitem [{\citenamefont {Abada}\ \emph {et~al.}(2019)\citenamefont {Abada},
  \citenamefont {Hern\'andez-Cabezudo},\ and\ \citenamefont
  {Marcano}}]{Abada:2018qok}%
  \BibitemOpen
  \bibfield  {author} {\bibinfo {author} {\bibfnamefont {A.}~\bibnamefont
  {Abada}}, \bibinfo {author} {\bibfnamefont {A.}~\bibnamefont
  {Hern\'andez-Cabezudo}}, \ and\ \bibinfo {author} {\bibfnamefont
  {X.}~\bibnamefont {Marcano}},\ }\href {\doibase 10.1007/JHEP01(2019)041}
  {\bibfield  {journal} {\bibinfo  {journal} {JHEP}\ }\textbf {\bibinfo
  {volume} {01}},\ \bibinfo {pages} {041} (\bibinfo {year} {2019})},\ \Eprint
  {http://arxiv.org/abs/1807.01331} {arXiv:1807.01331 [hep-ph]} \BibitemShut
  {NoStop}%
\bibitem [{\citenamefont {Dekens}\ \emph {et~al.}(2020)\citenamefont {Dekens},
  \citenamefont {de~Vries}, \citenamefont {Fuyuto}, \citenamefont
  {Mereghetti},\ and\ \citenamefont {Zhou}}]{Dekens:2020ttz}%
  \BibitemOpen
  \bibfield  {author} {\bibinfo {author} {\bibfnamefont {W.}~\bibnamefont
  {Dekens}}, \bibinfo {author} {\bibfnamefont {J.}~\bibnamefont {de~Vries}},
  \bibinfo {author} {\bibfnamefont {K.}~\bibnamefont {Fuyuto}}, \bibinfo
  {author} {\bibfnamefont {E.}~\bibnamefont {Mereghetti}}, \ and\ \bibinfo
  {author} {\bibfnamefont {G.}~\bibnamefont {Zhou}},\ }\href {\doibase
  10.1007/JHEP06(2020)097} {\bibfield  {journal} {\bibinfo  {journal} {JHEP}\
  }\textbf {\bibinfo {volume} {06}},\ \bibinfo {pages} {097} (\bibinfo {year}
  {2020})},\ \Eprint {http://arxiv.org/abs/2002.07182} {arXiv:2002.07182
  [hep-ph]} \BibitemShut {NoStop}%
\bibitem [{\citenamefont {Jedamzik}(2006)}]{Jedamzik:2006xz}%
  \BibitemOpen
  \bibfield  {author} {\bibinfo {author} {\bibfnamefont {K.}~\bibnamefont
  {Jedamzik}},\ }\href {\doibase 10.1103/PhysRevD.74.103509} {\bibfield
  {journal} {\bibinfo  {journal} {Phys. Rev. D}\ }\textbf {\bibinfo {volume}
  {74}},\ \bibinfo {pages} {103509} (\bibinfo {year} {2006})},\ \Eprint
  {http://arxiv.org/abs/hep-ph/0604251} {arXiv:hep-ph/0604251} \BibitemShut
  {NoStop}%
\bibitem [{\citenamefont {Kawasaki}\ \emph {et~al.}(2005)\citenamefont
  {Kawasaki}, \citenamefont {Kohri},\ and\ \citenamefont
  {Moroi}}]{Kawasaki:2004qu}%
  \BibitemOpen
  \bibfield  {author} {\bibinfo {author} {\bibfnamefont {M.}~\bibnamefont
  {Kawasaki}}, \bibinfo {author} {\bibfnamefont {K.}~\bibnamefont {Kohri}}, \
  and\ \bibinfo {author} {\bibfnamefont {T.}~\bibnamefont {Moroi}},\ }\href
  {\doibase 10.1103/PhysRevD.71.083502} {\bibfield  {journal} {\bibinfo
  {journal} {Phys. Rev. D}\ }\textbf {\bibinfo {volume} {71}},\ \bibinfo
  {pages} {083502} (\bibinfo {year} {2005})},\ \Eprint
  {http://arxiv.org/abs/astro-ph/0408426} {arXiv:astro-ph/0408426} \BibitemShut
  {NoStop}%
\bibitem [{\citenamefont {Burgess}\ \emph {et~al.}(2001)\citenamefont
  {Burgess}, \citenamefont {Pospelov},\ and\ \citenamefont {ter
  Veldhuis}}]{Burgess:2000yq}%
  \BibitemOpen
  \bibfield  {author} {\bibinfo {author} {\bibfnamefont {C.~P.}\ \bibnamefont
  {Burgess}}, \bibinfo {author} {\bibfnamefont {M.}~\bibnamefont {Pospelov}}, \
  and\ \bibinfo {author} {\bibfnamefont {T.}~\bibnamefont {ter Veldhuis}},\
  }\href {\doibase 10.1016/S0550-3213(01)00513-2} {\bibfield  {journal}
  {\bibinfo  {journal} {Nucl. Phys. B}\ }\textbf {\bibinfo {volume} {619}},\
  \bibinfo {pages} {709} (\bibinfo {year} {2001})},\ \Eprint
  {http://arxiv.org/abs/hep-ph/0011335} {arXiv:hep-ph/0011335} \BibitemShut
  {NoStop}%
\bibitem [{\citenamefont {Matos}\ and\ \citenamefont
  {Lopez-Fernandez}(2014)}]{Matos:2014gha}%
  \BibitemOpen
  \bibfield  {author} {\bibinfo {author} {\bibfnamefont {T.}~\bibnamefont
  {Matos}}\ and\ \bibinfo {author} {\bibfnamefont {R.}~\bibnamefont
  {Lopez-Fernandez}},\ }\href@noop {} {\  (\bibinfo {year} {2014})},\ \Eprint
  {http://arxiv.org/abs/1403.5243} {arXiv:1403.5243 [gr-qc]} \BibitemShut
  {NoStop}%
\bibitem [{\citenamefont {Wang}\ \emph {et~al.}(2022)\citenamefont {Wang},
  \citenamefont {Wu}, \citenamefont {Wu},\ and\ \citenamefont
  {Zhu}}]{Wang:2021oha}%
  \BibitemOpen
  \bibfield  {author} {\bibinfo {author} {\bibfnamefont {W.}~\bibnamefont
  {Wang}}, \bibinfo {author} {\bibfnamefont {K.-Y.}\ \bibnamefont {Wu}},
  \bibinfo {author} {\bibfnamefont {L.}~\bibnamefont {Wu}}, \ and\ \bibinfo
  {author} {\bibfnamefont {B.}~\bibnamefont {Zhu}},\ }\href {\doibase
  10.1016/j.nuclphysb.2022.115907} {\bibfield  {journal} {\bibinfo  {journal}
  {Nucl. Phys. B}\ }\textbf {\bibinfo {volume} {983}},\ \bibinfo {pages}
  {115907} (\bibinfo {year} {2022})},\ \Eprint
  {http://arxiv.org/abs/2112.06492} {arXiv:2112.06492 [hep-ph]} \BibitemShut
  {NoStop}%
\bibitem [{\citenamefont {Mancuso}\ \emph {et~al.}(2020)\citenamefont {Mancuso}
  \emph {et~al.}}]{CRESST:2020wtj}%
  \BibitemOpen
  \bibfield  {author} {\bibinfo {author} {\bibfnamefont {M.}~\bibnamefont
  {Mancuso}} \emph {et~al.} (\bibinfo {collaboration} {CRESST}),\ }\href
  {\doibase 10.1007/s10909-020-02343-3} {\bibfield  {journal} {\bibinfo
  {journal} {J. Low Temp. Phys.}\ }\textbf {\bibinfo {volume} {199}},\ \bibinfo
  {pages} {547} (\bibinfo {year} {2020})}\BibitemShut {NoStop}%
\bibitem [{\citenamefont {Aprile}\ \emph {et~al.}(2013)\citenamefont {Aprile}
  \emph {et~al.}}]{XENON100:2013ele}%
  \BibitemOpen
  \bibfield  {author} {\bibinfo {author} {\bibfnamefont {E.}~\bibnamefont
  {Aprile}} \emph {et~al.} (\bibinfo {collaboration} {XENON100}),\ }\href
  {\doibase 10.1103/PhysRevLett.111.021301} {\bibfield  {journal} {\bibinfo
  {journal} {Phys. Rev. Lett.}\ }\textbf {\bibinfo {volume} {111}},\ \bibinfo
  {pages} {021301} (\bibinfo {year} {2013})},\ \Eprint
  {http://arxiv.org/abs/1301.6620} {arXiv:1301.6620 [astro-ph.CO]} \BibitemShut
  {NoStop}%
\bibitem [{\citenamefont {Agrawal}\ \emph {et~al.}(2021)\citenamefont {Agrawal}
  \emph {et~al.}}]{Agrawal:2021dbo}%
  \BibitemOpen
  \bibfield  {author} {\bibinfo {author} {\bibfnamefont {P.}~\bibnamefont
  {Agrawal}} \emph {et~al.},\ }\href {\doibase 10.1140/epjc/s10052-021-09703-7}
  {\bibfield  {journal} {\bibinfo  {journal} {Eur. Phys. J. C}\ }\textbf
  {\bibinfo {volume} {81}},\ \bibinfo {pages} {1015} (\bibinfo {year}
  {2021})},\ \Eprint {http://arxiv.org/abs/2102.12143} {arXiv:2102.12143
  [hep-ph]} \BibitemShut {NoStop}%
\bibitem [{\citenamefont {Adamson}\ \emph {et~al.}(2011)\citenamefont {Adamson}
  \emph {et~al.}}]{MINOS:2011amj}%
  \BibitemOpen
  \bibfield  {author} {\bibinfo {author} {\bibfnamefont {P.}~\bibnamefont
  {Adamson}} \emph {et~al.} (\bibinfo {collaboration} {MINOS}),\ }\href
  {\doibase 10.1103/PhysRevLett.107.181802} {\bibfield  {journal} {\bibinfo
  {journal} {Phys. Rev. Lett.}\ }\textbf {\bibinfo {volume} {107}},\ \bibinfo
  {pages} {181802} (\bibinfo {year} {2011})},\ \Eprint
  {http://arxiv.org/abs/1108.0015} {arXiv:1108.0015 [hep-ex]} \BibitemShut
  {NoStop}%
\bibitem [{\citenamefont {Aguilar}\ \emph {et~al.}(2001)\citenamefont {Aguilar}
  \emph {et~al.}}]{LSND:2001aii}%
  \BibitemOpen
  \bibfield  {author} {\bibinfo {author} {\bibfnamefont {A.}~\bibnamefont
  {Aguilar}} \emph {et~al.} (\bibinfo {collaboration} {LSND}),\ }\href
  {\doibase 10.1103/PhysRevD.64.112007} {\bibfield  {journal} {\bibinfo
  {journal} {Phys. Rev. D}\ }\textbf {\bibinfo {volume} {64}},\ \bibinfo
  {pages} {112007} (\bibinfo {year} {2001})},\ \Eprint
  {http://arxiv.org/abs/hep-ex/0104049} {arXiv:hep-ex/0104049} \BibitemShut
  {NoStop}%
\bibitem [{\citenamefont {Abazajian}\ \emph {et~al.}(2012)\citenamefont
  {Abazajian} \emph {et~al.}}]{Abazajian:2012ys}%
  \BibitemOpen
  \bibfield  {author} {\bibinfo {author} {\bibfnamefont {K.~N.}\ \bibnamefont
  {Abazajian}} \emph {et~al.},\ }\href@noop {} {\  (\bibinfo {year} {2012})},\
  \Eprint {http://arxiv.org/abs/1204.5379} {arXiv:1204.5379 [hep-ph]}
  \BibitemShut {NoStop}%
\bibitem [{\citenamefont {Deppisch}\ \emph {et~al.}(2015)\citenamefont
  {Deppisch}, \citenamefont {Bhupal~Dev},\ and\ \citenamefont
  {Pilaftsis}}]{Deppisch:2015qwa}%
  \BibitemOpen
  \bibfield  {author} {\bibinfo {author} {\bibfnamefont {F.~F.}\ \bibnamefont
  {Deppisch}}, \bibinfo {author} {\bibfnamefont {P.~S.}\ \bibnamefont
  {Bhupal~Dev}}, \ and\ \bibinfo {author} {\bibfnamefont {A.}~\bibnamefont
  {Pilaftsis}},\ }\href {\doibase 10.1088/1367-2630/17/7/075019} {\bibfield
  {journal} {\bibinfo  {journal} {New J. Phys.}\ }\textbf {\bibinfo {volume}
  {17}},\ \bibinfo {pages} {075019} (\bibinfo {year} {2015})},\ \Eprint
  {http://arxiv.org/abs/1502.06541} {arXiv:1502.06541 [hep-ph]} \BibitemShut
  {NoStop}%
\bibitem [{\citenamefont {Dasgupta}\ and\ \citenamefont
  {Kopp}(2021)}]{Dasgupta:2021ies}%
  \BibitemOpen
  \bibfield  {author} {\bibinfo {author} {\bibfnamefont {B.}~\bibnamefont
  {Dasgupta}}\ and\ \bibinfo {author} {\bibfnamefont {J.}~\bibnamefont
  {Kopp}},\ }\href {\doibase 10.1016/j.physrep.2021.06.002} {\bibfield
  {journal} {\bibinfo  {journal} {Phys. Rept.}\ }\textbf {\bibinfo {volume}
  {928}},\ \bibinfo {pages} {1} (\bibinfo {year} {2021})},\ \Eprint
  {http://arxiv.org/abs/2106.05913} {arXiv:2106.05913 [hep-ph]} \BibitemShut
  {NoStop}%
\bibitem [{\citenamefont {Beacham}\ \emph {et~al.}(2020)\citenamefont {Beacham}
  \emph {et~al.}}]{Beacham:2019nyx}%
  \BibitemOpen
  \bibfield  {author} {\bibinfo {author} {\bibfnamefont {J.}~\bibnamefont
  {Beacham}} \emph {et~al.},\ }\href {\doibase 10.1088/1361-6471/ab4cd2}
  {\bibfield  {journal} {\bibinfo  {journal} {J. Phys. G}\ }\textbf {\bibinfo
  {volume} {47}},\ \bibinfo {pages} {010501} (\bibinfo {year} {2020})},\
  \Eprint {http://arxiv.org/abs/1901.09966} {arXiv:1901.09966 [hep-ex]}
  \BibitemShut {NoStop}%
\bibitem [{\citenamefont {Bandyopadhyay}\ \emph {et~al.}(2018)\citenamefont
  {Bandyopadhyay}, \citenamefont {Chun},\ and\ \citenamefont
  {Mandal}}]{Bandyopadhyay:2017bgh}%
  \BibitemOpen
  \bibfield  {author} {\bibinfo {author} {\bibfnamefont {P.}~\bibnamefont
  {Bandyopadhyay}}, \bibinfo {author} {\bibfnamefont {E.~J.}\ \bibnamefont
  {Chun}}, \ and\ \bibinfo {author} {\bibfnamefont {R.}~\bibnamefont
  {Mandal}},\ }\href {\doibase 10.1103/PhysRevD.97.015001} {\bibfield
  {journal} {\bibinfo  {journal} {Phys. Rev. D}\ }\textbf {\bibinfo {volume}
  {97}},\ \bibinfo {pages} {015001} (\bibinfo {year} {2018})},\ \Eprint
  {http://arxiv.org/abs/1707.00874} {arXiv:1707.00874 [hep-ph]} \BibitemShut
  {NoStop}%
\bibitem [{\citenamefont {Borah}\ \emph {et~al.}(2022)\citenamefont {Borah},
  \citenamefont {Dutta}, \citenamefont {Mahapatra},\ and\ \citenamefont
  {Sahu}}]{Borah:2021pet}%
  \BibitemOpen
  \bibfield  {author} {\bibinfo {author} {\bibfnamefont {D.}~\bibnamefont
  {Borah}}, \bibinfo {author} {\bibfnamefont {M.}~\bibnamefont {Dutta}},
  \bibinfo {author} {\bibfnamefont {S.}~\bibnamefont {Mahapatra}}, \ and\
  \bibinfo {author} {\bibfnamefont {N.}~\bibnamefont {Sahu}},\ }\href {\doibase
  10.1103/PhysRevD.105.015004} {\bibfield  {journal} {\bibinfo  {journal}
  {Phys. Rev. D}\ }\textbf {\bibinfo {volume} {105}},\ \bibinfo {pages}
  {015004} (\bibinfo {year} {2022})},\ \Eprint
  {http://arxiv.org/abs/2110.00021} {arXiv:2110.00021 [hep-ph]} \BibitemShut
  {NoStop}%
\end{thebibliography}%
\end{document}